# Complete long-term corrosion protection with chemical vapor deposited graphene

Feng Yu [a,*], Luca Camilli [a], Ting Wang [a], David M. A. Mackenzie [a], Michele Curioni [b], Robert Akid [b], Peter Bøggild [a,*]

[a] *CNG – DTU Nanotech, Department of Micro- and Nanotechnology, Technical University of Denmark, Kgs. Lyngby, DK-2800, Denmark*

[b] *School of Materials, The University of Manchester, Manchester, M13 9PL, UK*

**ABSTRACT:** Despite the numerous reports on the topic, examples of chemical vapor deposited (CVD) graphene-based anticorrosive coatings able to provide long-term protection (i.e., several months) of metals are still unavailable. Here, we finally present a polymer-graphene hybrid coating, comprising two single layers of CVD graphene sandwiched by three layers of polyvinyl butyral, that provides complete corrosion protection to commercial aluminum alloys even after 120 days of exposure to simulated seawater. The essential role played by graphene in the hybrid coating is evident when we compare the results from a polymer-only coating of the same thickness, which fails in protecting the metal after barely 30 days. With the emergence of commercially available large-area CVD graphene, our work demonstrates a straightforward approach towards high-performance anticorrosive coatings, which can be extended to other two-dimensional materials and polymers, for long-term protection of various relevant metals and alloys.

*Corresponding author. E-mail: fengy@nanotech.dtu.dk (Feng Yu),
Peter.Boggild@nanotech.dtu.dk (Peter Bøggild)

1. Introduction

The reliability and long-term durability of metal components is critical in many industrial sectors, such as aerospace, marine, transportation, construction, energy and manufacturing. Corrosion protection of metals is therefore vital to ensure useful component and system lifetimes, therein preventing economic loss, corrosion-induced catastrophic disasters and reducing negative impact on the environment. A well-established strategy for corrosion protection is to apply coatings on metal surfaces. Such anticorrosive coatings can consist of several layers, each with a specific function.[1, 2] Barrier layers, in particular, serve to separate metals from the environment. Graphene, being highly impermeable to gases[3] and chemically inert[4], has been considered a promising candidate as a physical barrier for corrosion protection,[5-9] following the seminal article by Ruoff's group.[4] While initial successes for graphene corrosion-inhibiting coatings have been reported,[4-7, 10] there are still several issues preventing the practical use of graphene in corrosion protection. For instance, high quality graphene is very challenging to grow directly on many commercially relevant metals and alloys (e.g., steel, Al and Mg alloys). Moreover, steels and other relevant alloys cannot in general withstand the high temperature needed for continuous graphene growth via CVD process. Alternatively, graphene coatings could be grown on a suitable growth substrate and subsequently transferred to a target metal surface. However, bare graphene coatings directly applied on metals show no or limited improvement in terms of corrosion protection[5, 11-13] due to weak adhesion of graphene on metals[14, 15], galvanic corrosion issues introduced by noble graphene [11, 12, 16, 17] and direct corrosion attack at sites where graphene defects located.[18-20] Using a graphene composite coating, where graphene-based nano-fillers (e.g. graphene oxide or reduced graphene oxide) are dispersed in a coating matrix is an alternative strategy.[21-27] Here, the fillers provide a tortuous diffusion pathway for corrosive species, thus enhancing the overall barrier performance of the coating. However, it is generally difficult to control the stacking of graphene-based nano-fillers in the coating matrix, and their agglomeration can ultimately limit the barrier performance of the coating by allowing in fact more diffusion pathways through the matrix.

Here we report on hybrid anticorrosive coatings consisting of alternating CVD single layer graphene (SLGr) and polymer films. A systematic study performed on coatings with a different number and combination of layers allows us to understand the specific function of each layer comprising the hybrid coating and identify whether the behavior and performance of such hybrid coatings is critically dependent on its overall structure. In this study, a commercial aluminum alloy(AA) AA2024-T3, is the substrate of choice as it is used extensively in the aerospace industry, due to its excellent strength-to-weight ratio.[28] We have found that the best performance is provided by coatings made up of two graphene layers sandwiched by three polymer layers. More specifically, when AA is coated with a polymer/SLGr/polymer/SLGr/polymer coating, its impedance at 0.01 Hz remains at $10^9$ $\Omega \cdot cm^2$ for up to 120 days of immersion in 3.5 wt% NaCl solution. Although several discouraging reports have previously questioned the suitability of graphene-based coatings for corrosion protection in practical, long-term applications,[11, 12] here, by combining the adhesive[29] and insulating properties of polymer layers with the impermeability of continuous CVD-grown graphene sheets, we demonstrate a facile and ultimately scalable method, which can be applied to any realistic metal substrate, to make coatings that exploit graphene's excellent barrier properties, while avoiding its intrinsic drawbacks, such as poor adhesion, galvanic corrosion and fast diffusion through defects.

## 2. Experimental

*2.1. Growth of graphene and preparation of polymer-graphene hybrid films*

A copper foil (25µm-thick, Part No. 13382, Alfa Aesar) was electrochemically polished in a solution with a volume ratio of 15% absolute ethanol (Millipore Corporation) and 85% concentrated phosphoric acid (Millipore Corporation) with magnetic stirring. A current density of ~0.04 A·cm$^{-2}$ was applied (Keithley 2400) on the copper foil for 3 minutes to reduce the surface roughness. The sample was then rinsed with deionized water and blow dried with nitrogen. The copper foil was loaded in a 4-inch graphite sample holder and thermal annealed in Ar (1000 sccm) at 1000 °C for 10 minutes at 25 mbar. The SLGr growth was carried out at atmospheric pressure for 15 minutes with a co-flow of Ar (900 sccm), H$_2$ (60 sccm) and CH$_4$ (2 sccm) in a commercial rapid thermal-CVD system (AS-ONE, Annealsys). After growth, samples were cooled down to room temperature at a rate of ~20 °C s$^{-1}$.

The polymer used to transfer graphene is a co-polymer of Polyvinyl butyral (PVB) (Mowital B 60 H, Kuraray Europe). The as-prepared graphene-covered copper foil was spin-coated with 12 wt% PVB ethanol solution at 1000 rpm for 1 minute and cured at 60 °C for 2 hours. The copper substrate was chemically etched in a mixed solution of 100 ml 5 wt% HCl (Millipore Corporation) and 3 ml 30 wt% H$_2$O$_2$ (Sigma-Aldrich). The polymer-graphene (P-G) films were then rinsed with deionized water. The preparation process of P-G film is illustrated in Fig. 1. The quality of as-grown and transferred graphene on SiO$_2$ or PVB was assessed by optical microscopy and Raman spectroscopy (Fig. S1-4, Supplementary Material).

*2.2. Fabrication of polymer-graphene hybrid coatings on aluminum alloy*

Firstly, AA (Wilsons metals, UK) was ultra-sonicated in acetone, de-smutted (1 minute at 60 °C) in 10 wt% NaOH (Sigma-Aldrich), etched (1 minute at 25 °C) in 50 wt% $HNO_3$ (Millipore Corporation), rinsed with deionized water and blow dried with $N_2$ to remove surface oxides. Then 12 wt% PVB ethanol solution was spin coated (1000 rpm, 1 minute) on AA and cured at 60 °C for 2 hours to coat the PVB primer on AA (AA-P). PVB is widely used for corrosion protection[28, 30] and is chosen here as it exhibits strong adhesion to the surface of AA.[29] The P-G film obtained as described above was then dried and transferred onto either AA or AA-P, using a polydimethylsiloxane (Dow Corning Corporation) stamp for the mechanical transfer. Thermal annealing was then applied at 100 °C for 10-15 minutes to prepare AA-P-G, AA-G-P and AA-P-G-P. Two layers of P-G film could also be transferred onto AA-P to prepare AA-P-G-P-G-P (or AA-P-P-G-G-P). Fig. 1 shows the fabrication procedures of these coatings. The reference samples for AA-P-G-P and AA-P-G-P-G-P (or AA-P-P-G-G-P) were prepared by transferring one or two layers of PVB, which were obtained after etching the PVB spin-coated copper, to AA-P and then thermal annealed. Such reference samples are denoted as AA-P-P and AA-P-P-P, respectively. Both the PVB primer directly spin-coated on AA and the PVB layer used for graphene transfer have a similar thickness, ranging between 4 µm and 5 µm (Fig. S6, Supplementary Material).

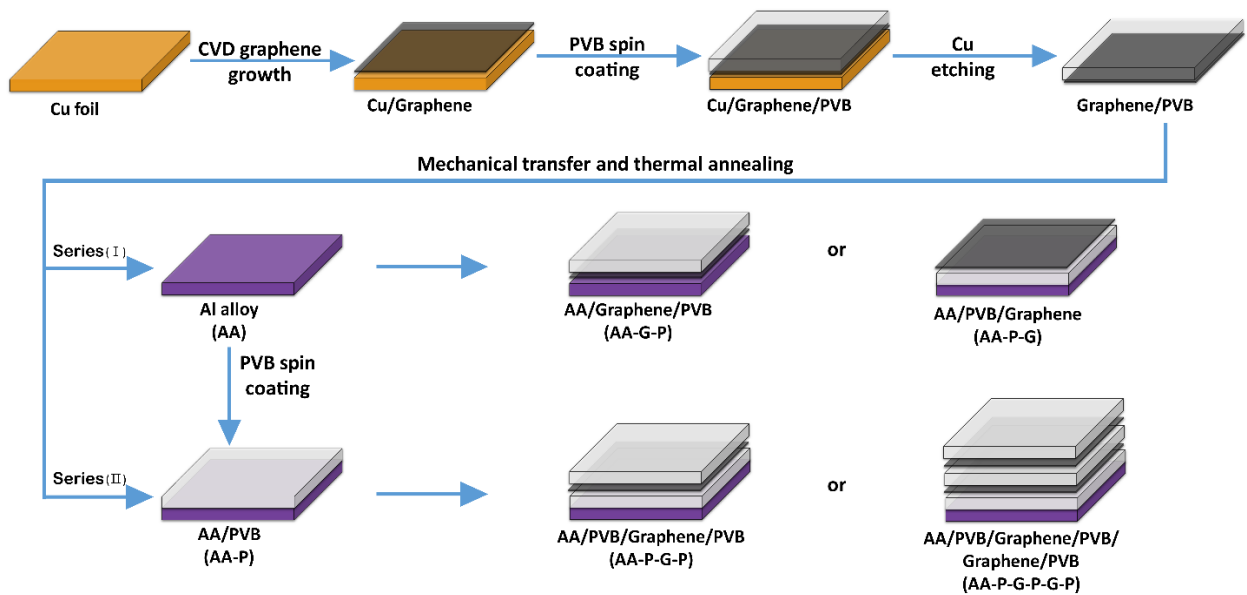

Fig. 1. Schematic illustration of preparation steps for the investigated polymer-graphene hybrid coatings on AA.

## 2.3. Electrochemical tests

All electrochemical measurements were conducted at room temperature in a custom-made three-electrode Teflon cell using the studied samples as the working electrode, a Ag/AgCl reference electrode, a platinum counter electrode, and 3.5 wt% NaCl solution (i.e., simulated seawater) as electrolyte. A Gamry Reference 3000 potentiostat and an ECM8 Electrochemical Multiplexer were used. Electrochemical impedance spectroscopy (EIS) were recorded at a frequency range from 100,000 Hz to 0,01 Hz with 10 points per decade under ±10 mV sinusoidal perturbation at the measured open circuit potential of the sample. Potentiodynamic scans (PDS) were performed with a scan rate of 1 mV/s starting from -500 mV (or -300 mV) to +500 mV (or +300 mV) vs OCP (or Ag/AgCl). Values of corrosion rate are calculated using Faradays' law from corrosion current density, which is obtained from Tafel analysis on PDS curves.

*2.4. Additional characterization techniques*

Optical images were collected using a Nikon Eclipse L200N optical microscope, whereas the SEM images were taken with a Quanta 200 FEG environmental scanning electron microscope. While optical images were collected on the samples without removing the coating, SEM images were obtained after all coatings were dissolved and removed with acetone. The Raman spectroscopy data were collected by a DXR Raman microscope (Thermo Fisher Scientific) using 455 nm laser with 8 mW power and 50X objective. Peaks were determined via Matlab as described by Larsen et al.[31]

**3. Results and discussion**

*3.1. Electrochemical testing of the coating performance*

Electrochemical impedance spectroscopy (EIS) is a powerful tool to evaluate the corrosion performance of coatings and provides insight into the corrosion behavior of coated metals as well as the barrier properties of anticorrosive coatings. A higher magnitude of impedance at low frequencies (e.g. $|Z|_{0.01Hz}$) is associated with increased anticorrosion performance, while a slow decrease of its value with elapsed time generally indicates high resistance to environmental degradation of the anticorrosive coatings.[32]

As the thickness of SLGr is negligible when compared to the thickness of the polymer layers, in our tests we consistently compare bare PVB layered reference coatings to their graphene-enhanced counterparts of nominally same thickness (for instance, we use AA-P-P as a reference for AA-P-G-P and so on).

Fig. 2a, b show respectively representative Bode and phase angle plots of both coated and uncoated samples after 1 day of immersion in 3.5 wt% NaCl solution (the complete set of data is provided in the Supplementary Material). The system AA-P-G-P, $|Z|_{0.01Hz}$ shows a value of $1.4\times10^7$ $\Omega\cdot cm^2$, i.e., 500 times higher than uncoated AA ($2.8\times10^4$ $\Omega\cdot cm^2$) and 20 times higher than the polymer-coated sample (AA-P-P) used as reference ($8.0\times10^5$ $\Omega\cdot cm^2$). This result proves the enhancement provided by a single graphene sheet (which is one-atom-thick) to the overall barrier properties of a ~10 µm thick polymer coating at short term. When the top polymer layer is absent (AA-P-G), we observe that $|Z|_{0.01Hz}$ decreases by an order of magnitude with respect to AA-P-G-P, ($|Z|_{0.01Hz}=1.8\times10^6$ $\Omega\cdot cm^2$), but is still 2 times higher than that of AA-P-P. Here it is worth pointing out that just replacing the 5µm-thick PVB top-layer with a one-atom thick graphene sheet leads to slightly better short-term anticorrosion performance. When the polymer primer is absent (AA-G-P), the value of low-frequency impedance becomes lower than the bare AA ($|Z|_{0.01Hz}=2.5\times10^4$ $\Omega\cdot cm^2$). The reduced performance of AA-G-P is attributed to galvanic coupling between the AA and graphene. This last result is in apparent disagreement with that reported by Yu et al.,[33] where a coating of CVD graphene covered with polymethyl methacrylate layer offered effective corrosion protection of a copper substrate. However, it is in agreement with recent findings[11, 12, 34] reporting that graphene layers directly in contact with metal substrates indeed accelerate the corrosion process due to the formation of a galvanic coupling between the noble and electron-conductive graphene sheet and the metal substrate. We therefore conclude that the insulating PVB, as it electrically separates graphene from the metal substrate, avoids the galvanic coupling.

Additionally, we test different coatings with two layers of graphene. An improvement of 50,000 times of the magnitude of $|Z|_{0.01Hz}$ with respect to uncoated AA is achieved by AA-P-G-P-

G-P ($1.4\times10^9$ $\Omega\cdot cm^2$) after 1 day of immersion. In contrast, the polymer-only reference sample of AA-P-P-P ($8.3\times10^7$ $\Omega\cdot cm^2$) showed only 3000 times improvement over that of uncoated AA after 1 day of immersion. Once again, the significant improvement to the overall barrier performance of the hybrid coating provided by graphene is demonstrated. EIS data recorded after short-term exposure to a corrosive agent may reflect the intrinsic barrier performance of coatings, but long-term tests are needed to evaluate the environmental degradation of the coating and explore its potential for real applications. Therefore, EIS tests are performed on AA, AA-P-G-P, AA-P-P-P, AA-P-G-P-G-P at 30 days (Fig. 2c, d) and on AA-P-G-P-G-P at 120 days (Fig. 2e, f).

After 30 days of immersion, the magnitude of $|Z|_{0.01Hz}$ for AA-P-G-P drops to 7% of the value (from $1.4\times10^7$ $\Omega\cdot cm^2$ to $9.2\times10^5$ $\Omega\cdot cm^2$), while the value for AA-P-P-P drops to 3% (from $8.3\times10^7$ $\Omega\cdot cm^2$ to $2.4\times10^6$ $\Omega\cdot cm^2$), suggesting that both coatings undergo severe degradation and therefore are not effective for long-term corrosion protection of AA. On the other hand, AA-P-G-P-G-P shows only a minimal decrease of the magnitude of $|Z|_{0.01Hz}$, which even after 120 days of immersion remains in the $10^9$ $\Omega\cdot cm^2$ range, indicating both excellent barrier performance and high resistance to environmental degradation.

We have also performed EIS measurements on some of our coatings applied to brass and steel. The data is reported in Supplementary Material (Fig. S13) and shows that our hybrid coating system can indeed provide effective protection when applied to other substrates as well. This clearly highlights the flexibility of the presented approach. A complete set of EIS data for all studied samples are presented in Supplementary Material (Fig. S8, S10, S12, S13, S15, S18).

From the phase spectra reported in Fig. 2b, d and f, one can notice that an additional time constant (evident as a peak in the medium frequency range in the phase spectrum and a plateau in the impedance modulus spectrum) is observed for AA-P-G-P and AA-P-G. This time constant appears to be associated with the presence of a graphene layer within the film, and it is possibly related to the presence of the graphene-polymer interface. The specimens with three layers of polymer, even when they contain graphene layers, do not show such an additional time constant since they are considerably more resistive than the others and therefore the additional time constant cannot be resolved.

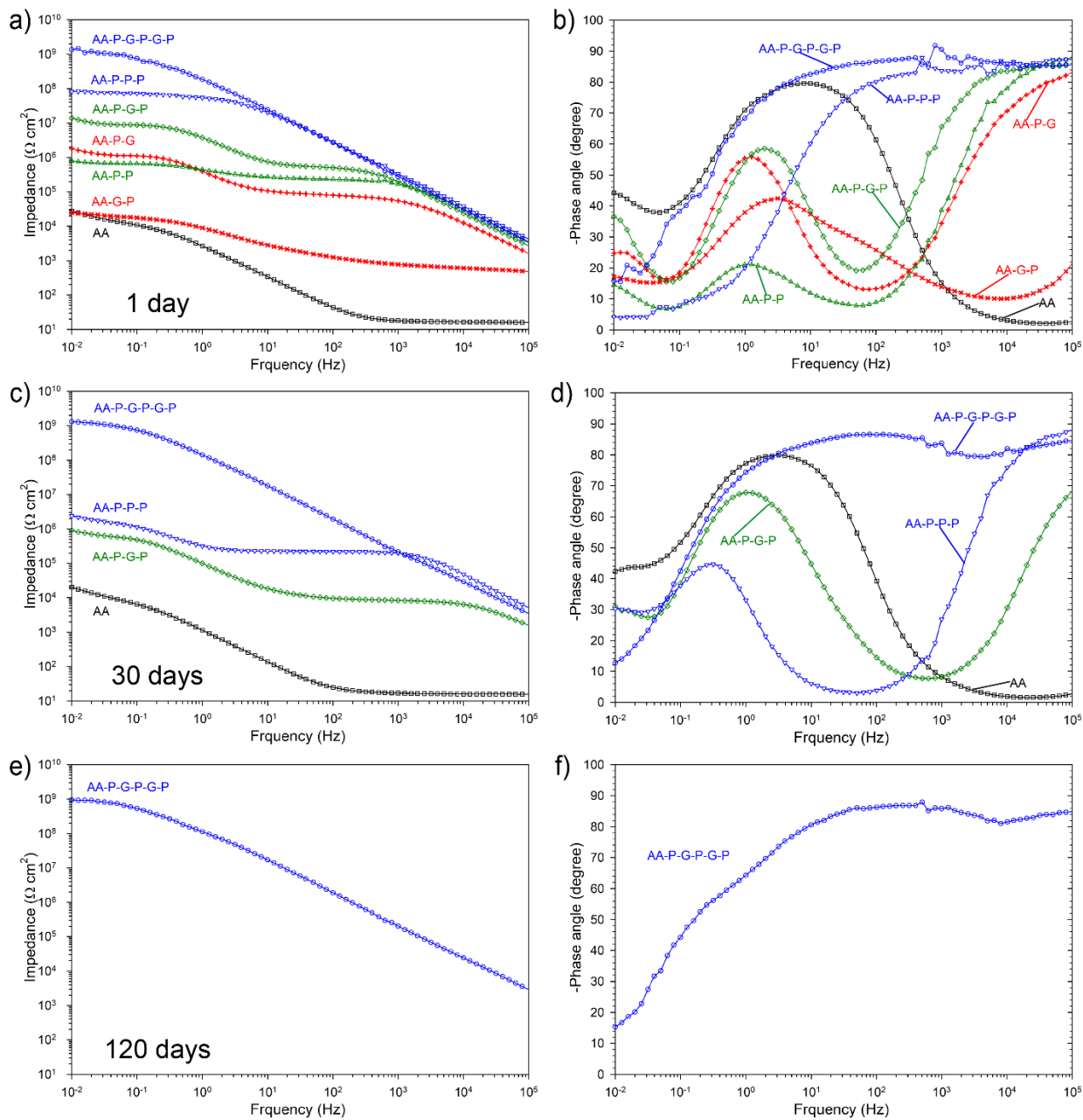

Fig. 2. Bode plots and phase diagram after 1 day (a and b, respectively), 30 days (c and d), and 120 days (e and f) of immersion in 3.5 wt% NaCl solution of tested samples. Note legends for each plot are identical: AA (black square), AA-G-P (red x), AA-P-G (red +), AA-P-P (green triangle up), AA-P-G-P (green diamond), AA-P-P-P (blue triangle down) and AA-P-G-P-G-P (blue circle).

Although potentiodynamic scans (PDS) have intrinsic limitations when used on coated metals and should be used with caution in such cases, they have been carried out on all samples, as this technique has often been used to test corrosion inhibition performance of graphene-coated samples.[5, 7, 24, 25]

Representative PDS measurements for the samples AA, AA-P-P-P and AA-P-G-P-G-P after 1, 30 or 120 days of immersion in 3.5 wt% NaCl solution are displayed in Fig. 3, while the complete set of data (for the other samples as well) is provided in the Supplementary Material. When AA is covered by P-P-P coatings, it exhibits a corrosion rate after 1 day of immersion that is two orders of magnitude lower than bare AA (20 nm/year for AA-P-P-P vs. 4 µm/year for bare AA). However, the corrosion rate of P-P-P coated AA increases by one order of magnitude after barely 30 days of immersion (0.3 µm/year for AA-P-P-P at 30 days vs. 20 nm/year for AA-P-P-P at 1 day), indicating the degradation of P-P-P coating. On the other hand, the PDS curves for AA-P-G-P-G-P remain in the relative low current range with the corrosion rate consistently below 2 nm/year over 120 days of immersion, suggesting no degradation of the coating within the tested timeframe. Overall, the PDS measurements are in agreement with the results from the EIS experiments.

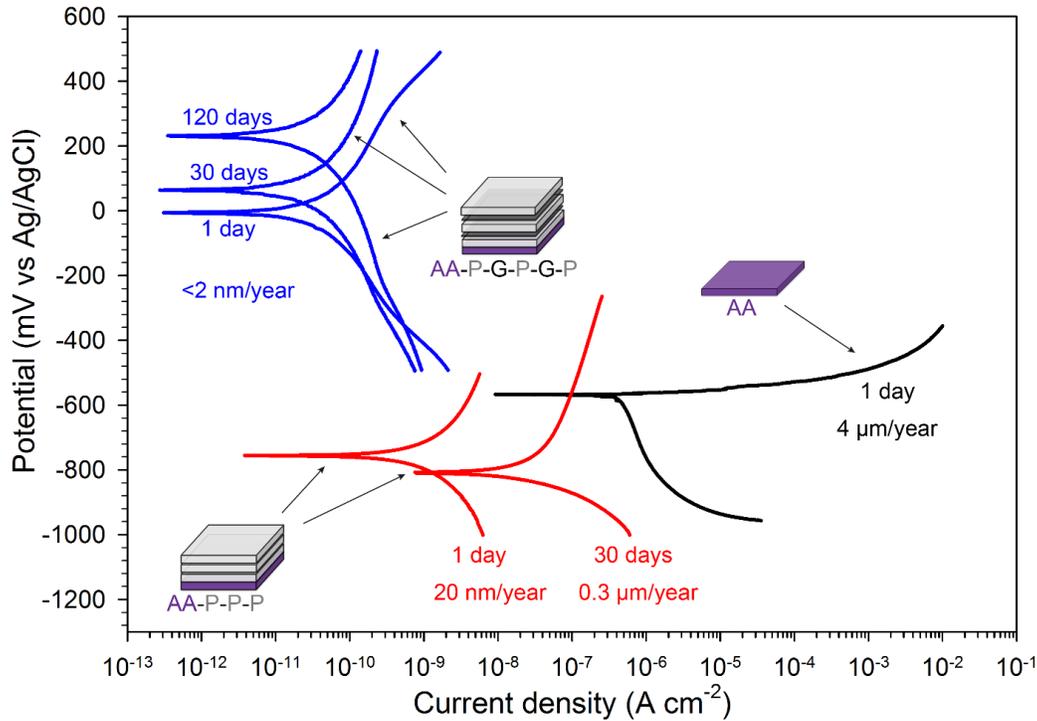

Fig. 3. Typical potentiodynamic scans for AA (black), AA-P-P-P (red) and AA-P-G-P-G-P (blue) after 1, 30 or 120 days of immersion in 3.5 wt% NaCl solution.

To better visualize our results, we plot the open circuit potential (OCP) and the corrosion current density vs low frequency impedance ($|Z|_{0.01Hz}$) for coated and uncoated samples in Fig. 4. Both OCP and impedance values are obtained from raw data of the electrochemical measurements, while corrosion current densities (corrosion rates) are obtained from PDS curves with Tafel analysis.

One realizes that we can cluster our data into three groups. Group 1 consists of AA and AA-G-P. Both samples have no polymer in direct contact with the AA (i.e., they do not have the polymer primer layer), and are characterized by small $|Z|_{0.01Hz}$, large corrosion current density and OCP in the range -600 mV to -800 mV.

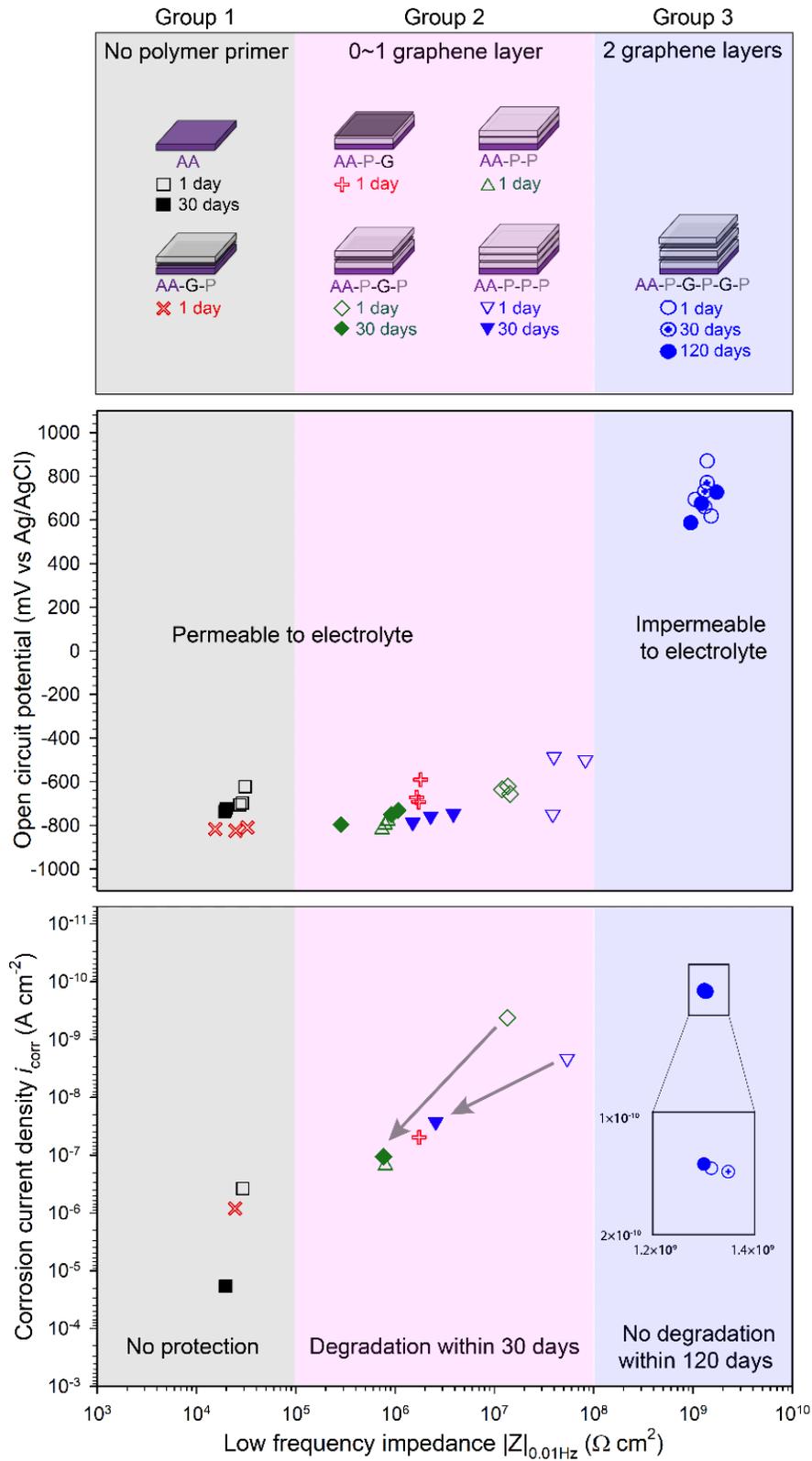

Fig. 4. Grouping of all tested samples and summary of the results from electrochemical tests.

In Group 2 we have polymer-only coatings and hybrid coatings with only one graphene layer, that is AA-P-G, AA-P-P, AA-P-G-P and AA-P-P-P. Although they show some differences, their overall behavior is similar. In particular, they do offer some protection, but they cannot prevent the electrolyte from attacking the substrate, leading to OCP values comparable to that of uncoated sample. Furthermore, the arrows in this group are indicating the degradation of AA-P-G-P and AA-P-P-P from 1 day to 30 days (empty *vs* filled symbols, respectively).

Group 3 consists of AA-P-G-P-G-P, which shows a behavior that is completely different from the samples in Group 1 and 2. Indeed, the P-G-P-G-P coating shows excellent barrier properties, with an OCP value in the range 600 mV to 900 mV, approximately 1.5 V higher than that of AA. In addition, AA-P-G-P-G-P exhibits the highest $|Z|_{0.01Hz}$ values and lowest corrosion current density, which, most importantly, remain almost unchanged within the 120 day-long experiment.

We want to emphasize that AA-P-G-P-G-P is the first CVD graphene-based coating that maintains high-performance in terms of $|Z|_{0.01Hz}$, corrosion current density and OCP over 120 days or, in other words, it is the first successful long-term anticorrosive coating based on CVD graphene thus far developed.

### 3.2. Characterization of corrosion morphology

Furthermore, we characterized all samples using optical and scanning electron microscopy (SEM), in order to give additional information about the nature of corrosion. Notably, the corrosion morphologies of AA-P-G-P, AA-P-P-P after 30 days and AA-P-G-P-G-P after both 30 days and 120 days of immersion in 3.5 wt% NaCl solution are displayed in Fig. 5. Both optical and SEM images of AA-P-G-P and AA-P-P-P coatings after 30 days of immersion show heavy corrosion of the two samples, in agreement with electrochemical measurements. In all reported examples, the areas shown represent the highest level of damage found on the sample.

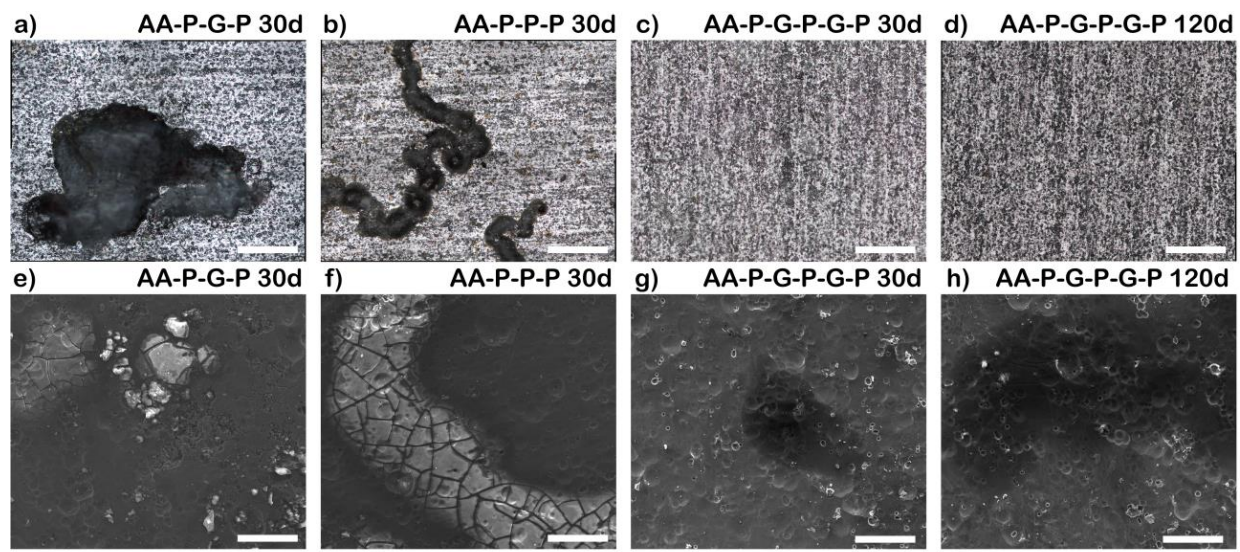

Fig. 5. Corrosion morphology of bare polymer and polymer-graphene hybrid coatings, where areas with most damage have been selected when possible a) – d) Optical and e) – h) SEM images of a), e) AA-P-G-P, b), f) AA-P-P-P and c), g) AA-P-G-P-G-P after 30 days and d), h) AA-P-G-P-G-P after 120 days of immersion in 3.5 wt% NaCl solution. Scale bars in a) – d) are 500μm and e) – h) are 50μm.

A complete set of images for all the samples is reported in Supplementary Material. While AA covered with P-G-P coating shows very mild corrosion at 1 day of immersion (Fig. S11b, e), the situation changes dramatically after 30 days of immersion, as concentrated pitting corrosion at millimeter scale is clearly observed. The corrosion for the sample AA-P-P-P develops on a large scale after 30 days of immersion, as seen from the filiform corrosion in Fig. 5b, f. Filiform corrosion is also observed for the sample AA-P-P at 1 day of immersion (Fig. S11a, d). For AA-

P-G-P-G-P coatings at 30 and 120 days of immersion, as presented in Fig. 5c, d, no visible corrosion can be observed from the optical images. Even from SEM examination, the surface of P-G-P-G-P coated AA after 30 and 120 days of immersion is homogeneous and smooth, similar to that of unexposed AA surface. In summary, the morphological investigation further confirms the limited protection performance of AA-P-P-P and AA-P-G-P samples for long-term exposure, while highlighting the excellent performance provided by AA-P-G-P-G-P.

*3.3. Corrosion protection mechanism*

Defect-free graphene is known to be impermeable to any molecule[3]. CVD graphene, however, naturally exhibits defects and tears through which molecules and other chemical species can easily pass through. When CVD graphene is applied to a metal substrate to protect it from environmental degradation, the corrosive and oxidizing agents will pass through graphene's defects and start the corrosion of the underlying substrate (Fig. 6a). Owing to graphene´s high electrical conductivity, the corrosion started locally under graphene's defects and will eventually spread throughout the whole metal substrate.[11, 12] On the other hand, although they are insulating, thicker and may offer good adhesion to metal substrates, polymer anticorrosive coatings are not as impermeable as graphene, therefore they are finally bound to fail over time (Fig. 6b). Hence, adding a continuous single sheet of graphene to a polymer coating greatly enhances its barrier properties, and provides effective corrosion protection at short term. Yet, over time, the corrosive species absorbed by the topmost polymer film will pass through graphene defects and tears, diffuse through the bottom polymer layer and eventually reach the metal surface, thus initiating its degradation (Fig. 6c). On the other hand, diffusion of corrosive species can be dramatically limited and retarded by adding two SLGr sheets to a polymer coating. (Such coatings can have either P-G-P-G-P or P-P-G-G-P structure, see Supplementary Material) The resulting coating indeed provides outstanding protection of aircraft aluminum for as long as four months of immersion in simulated seawater (Fig. 6d).

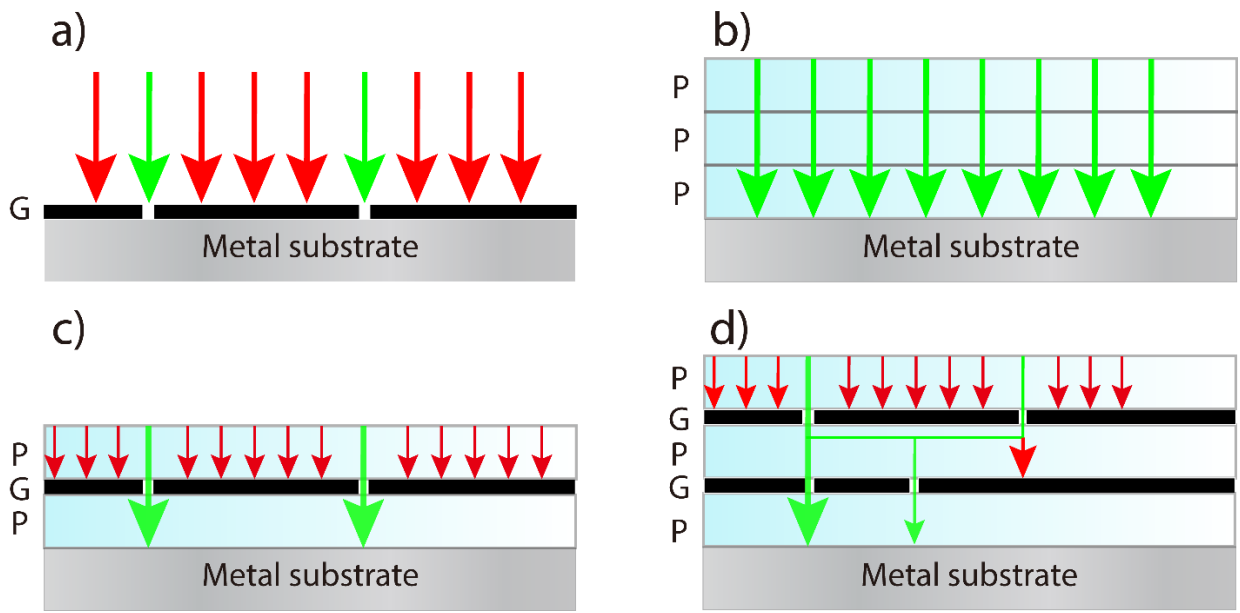

Fig. 6. Schematic illustration of the corrosion protection mechanism of a) as-grown graphene (G), b) bare polymer (P), c) P-G-P and d) P-G-P-G-P coatings on metal substrate.

## 4. Conclusion

In summary, we have prepared a polymer-graphene hybrid anticorrosive coating that optimally exploits the highly impermeable nature of graphene, and studied in detail the importance and function of both graphene as well as polymer layers within the hybrid coating. While a single layer CVD graphene between two polymer films (P-G-P) provides corrosion protection only for short-term (30 days), complete long-term (120 days) corrosion protection is achieved by sandwiching two single layers of CVD graphene between three polymer films (P-G-P-G-P or P-

P-G-G-P). It may be argued that our coatings are, in essence, polymer coatings, but the improvement of the graphene-containing coatings compared to the graphene-free variants, which were used as reference, highlights the substantial contribution provided by graphene, which turns out to make a crucial difference after 120 days of exposure to 3.5 wt% NaCl solution.

Finally, by reporting for the first time effective long-term (i.e., four months) protection of anticorrosive coatings based on CVD graphene, our findings may pave the way for the application of CVD graphene in the field of corrosion protection. In particular, since CVD graphene can be prepared via roll-to-roll processes, and roll-based lamination or processing has been demonstrated to work in practice,[35] we anticipate that this type of polymer-graphene hybrid could provide a high performance coating that can be applied as a dry foil to many different surfaces.

## Acknowledgements

This project was supported by The Strategic Danish Research collaboration (DSF) with the DA-GATE project (12-131827) and through the high technology foundation (HTF) with the NIAGRA project (058-2012-4), with additional support from the Danish National Research Foundation (DNRF), Center for Nanostructured Graphene (DNRF103) and the EU Graphene Flagship (604391). L.C. acknowledges support from the European Union's Horizon 2020 research and innovation program under the Marie Sklodowska-Curie grant agreement No. 658327. This project has received funding from the EU Horizon 2020 research and innovation program under grant agreement No. 696656.

## Supplementary Material

Characterization of graphene quality and coating thickness, optical microscopy and SEM study of corrosion morphology, complete set of polarization curves and electrochemical impedance spectroscopy of all tested samples, list of corrosion performance on reported results and this work. This material is available in the online version.

*Supplementary Material for*

# Complete long-term corrosion protection with chemical vapor deposited graphene

Feng Yu [a,*], Luca Camilli [a], Ting Wang [a], David M. A. Mackenzie [a], Michele Curioni [b], Robert Akid [b], Peter Bøggild [a,*]

[a] *CNG – DTU Nanotech, Department of Micro- and Nanotechnology, Technical University of Denmark, Kgs. Lyngby, DK-2800, Denmark*

[b] *School of Materials, The University of Manchester, Manchester, M13 9PL, UK*

*Address correspondence to fengy@nanotech.dtu.dk, Peter.Boggild@nanotech.dtu.dk.

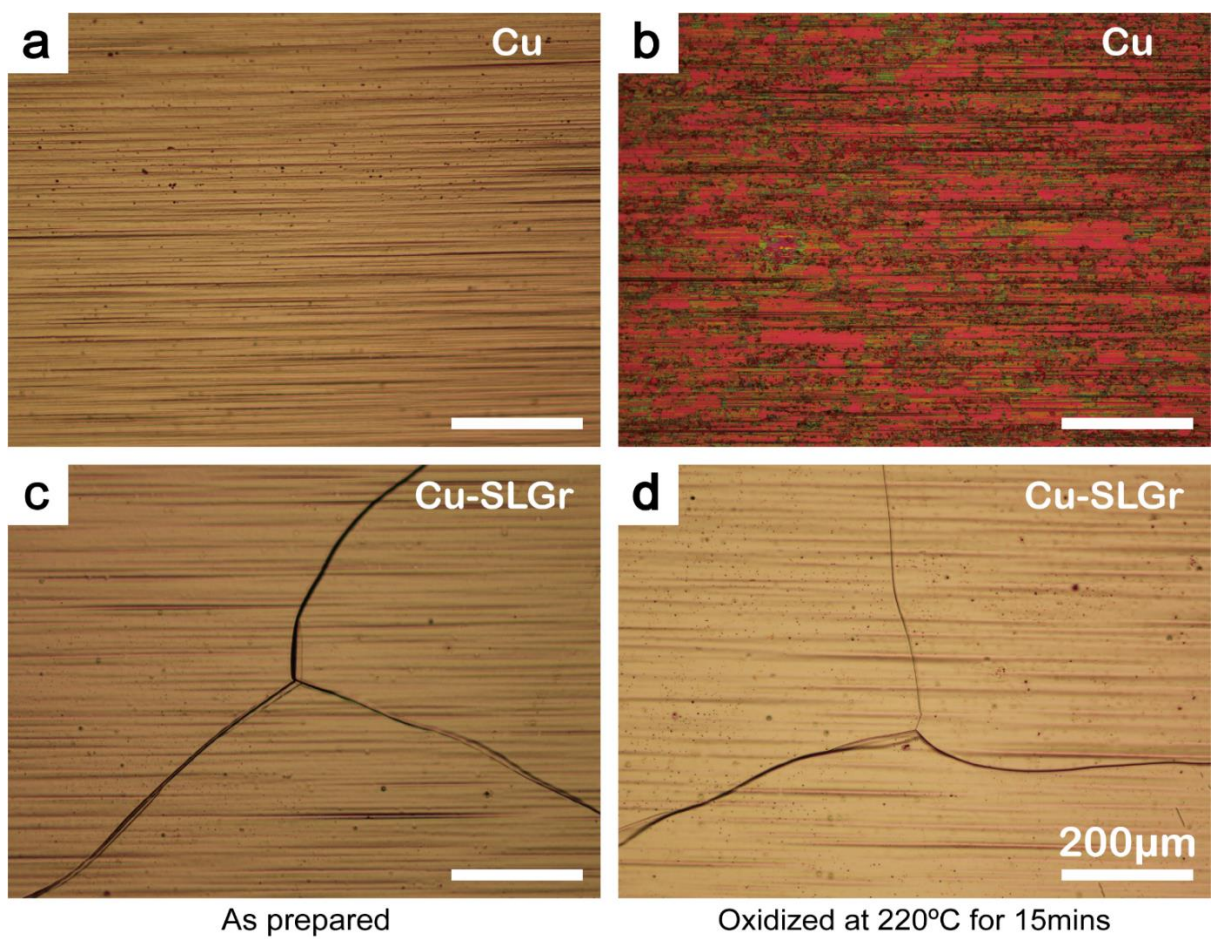

Fig. S1. Optical images of (a) fresh bare electrochemically polished copper (Cu), (b) Cu after oxidation in air at 200 °C for 15 minutes, (c) as prepared graphene covered copper (Cu-SLGr) and (d) Cu-SLGr after oxidation in air at 200 °C for 15 minutes.

After oxidation, bare Cu was severely oxidized to $Cu_2O$ (red color in b), while SLGr covered Cu just showed minimal oxidation (microscale red dots in d), indicating both the excellent barrier properties of SLGr and its full coverage on Cu. Note that the horizontal trench lines are the rolling lines on Cu surface during the preparation of the foils and the intersecting lines in c, d are grain boundaries of Cu formed during the high temperature annealing process.

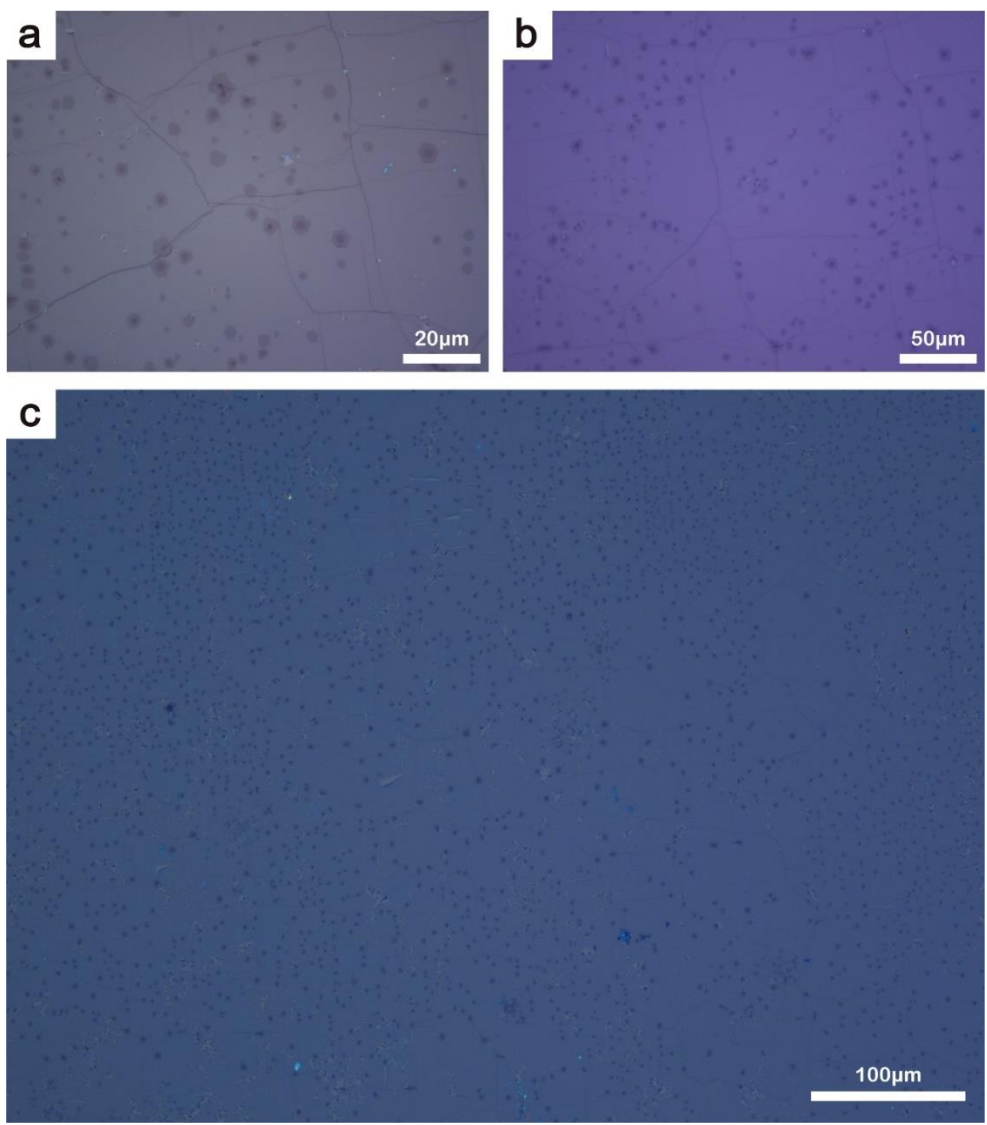

Fig. S2. Optical images of PVB transferred SLGr onto 90 nm thick $SiO_2$ wafer at (a) high, (b) medium and (c) low magnification.

CVD SLGr was transferred with PVB from a copper substrate to a $SiO_2$ wafer. However, a few micron sized pinholes (lighter areas in a, b, and c) were found on the transferred graphene layer, which may have originated from either the CVD growth process or the transfer process. Moreover, darker areas represent seeds of second or third graphene layers.

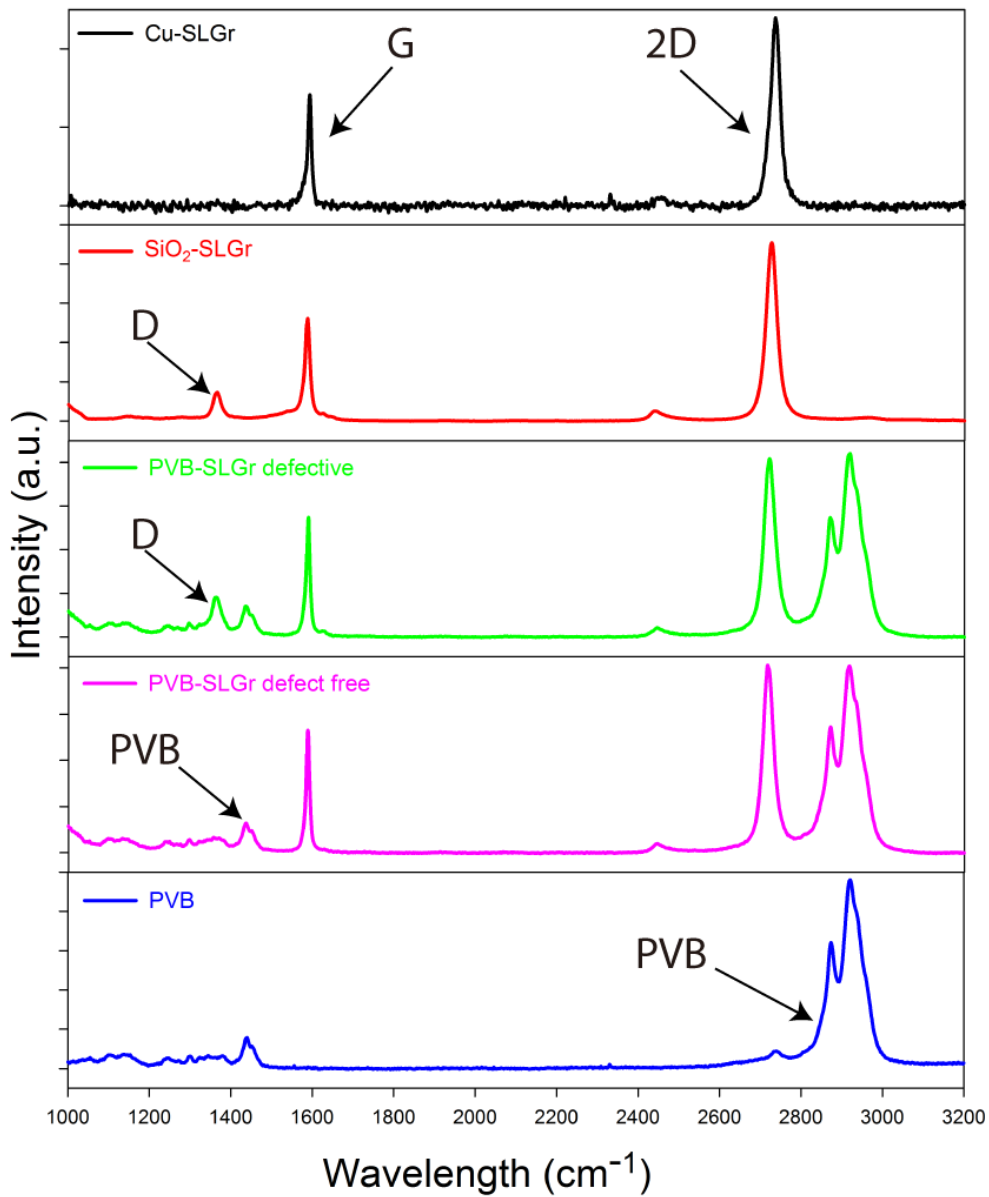

Fig. S3. Raman spectra of as-grown SLGr on copper (black), SLGr transferred onto SiO$_2$ (red), PVB supported SLGr with defects (green) and without defects (pink) and bare PVB (blue).

G (~1600 cm$^{-1}$) and 2D (~2750 cm$^{-1}$) peaks of graphene are presented regardless of the supporting substrate of copper, SiO$_2$ or PVB. However, defects from graphene are observed when supported on PVB or transferred onto SiO$_2$, as seen from the D peak (~1375cm$^{-1}$). The peaks between 2800 and 3000 cm$^{-1}$ and between 1430 and 1450 cm$^{-1}$ can be attributed to PVB.

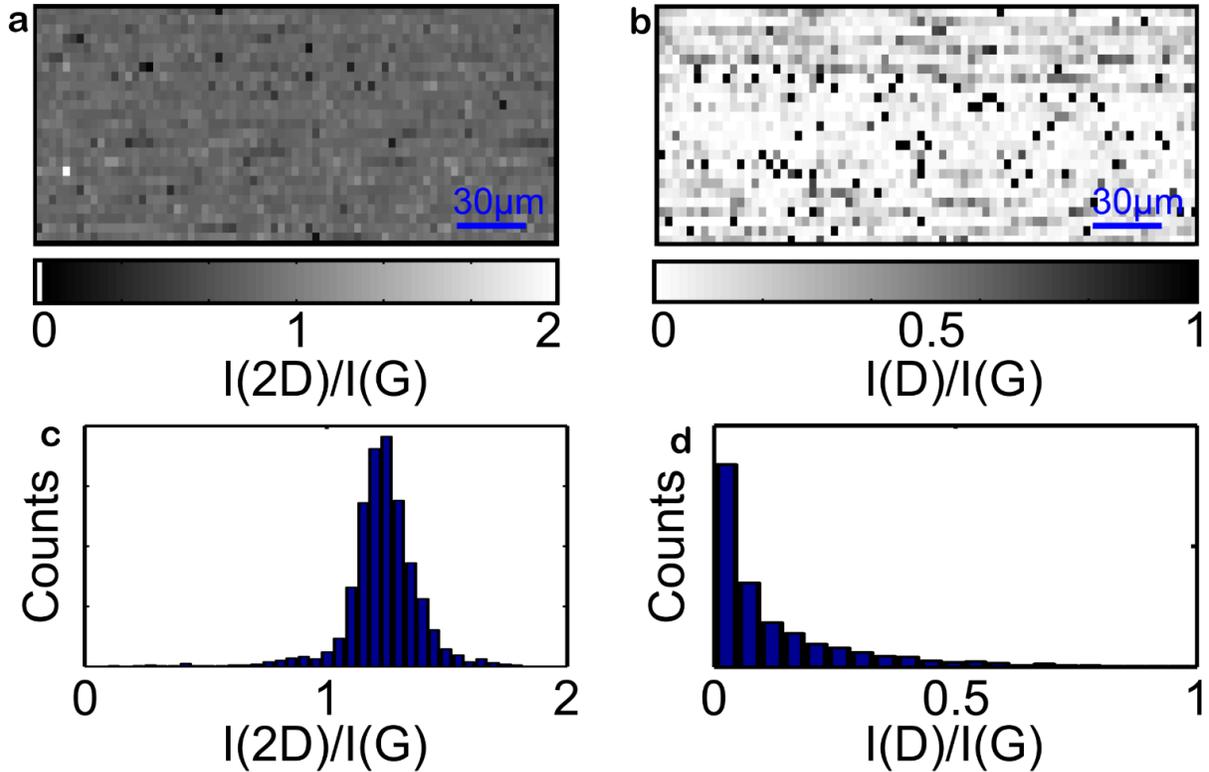

Fig. S4. Raman spectroscopic maps of the (a) I(2D)/I(G) and (b) I(D)/I(G) peak ratios of PVB supported graphene films on SiO$_2$ substrate. Statistical distribution of (c) I(2D)/I(G) and (d) I(D)/I(G) peak ratios.

The graphene/PVB layer is directly transferred to SiO$_2$ with the graphene layer facing up and then subject to thermal annealing. Micro-Raman spectroscopy study is directly carried out on the graphene layer in a representative 168 µm×102 µm area (a, b). The mapping is conducted with a step size of 3 µm using a 455 nm laser. Graphene is successfully transferred to PVB layer with a coverage of ~99%. The transferred graphene layer can be defective, as seen from the dark points in (b), with 8.1% spectra having a value of I(D)/I(G) higher than 0.5 (d). Moreover, we have demonstrated, for the first time, that the co-polymer PVB can be used as a graphene transfer support layer.

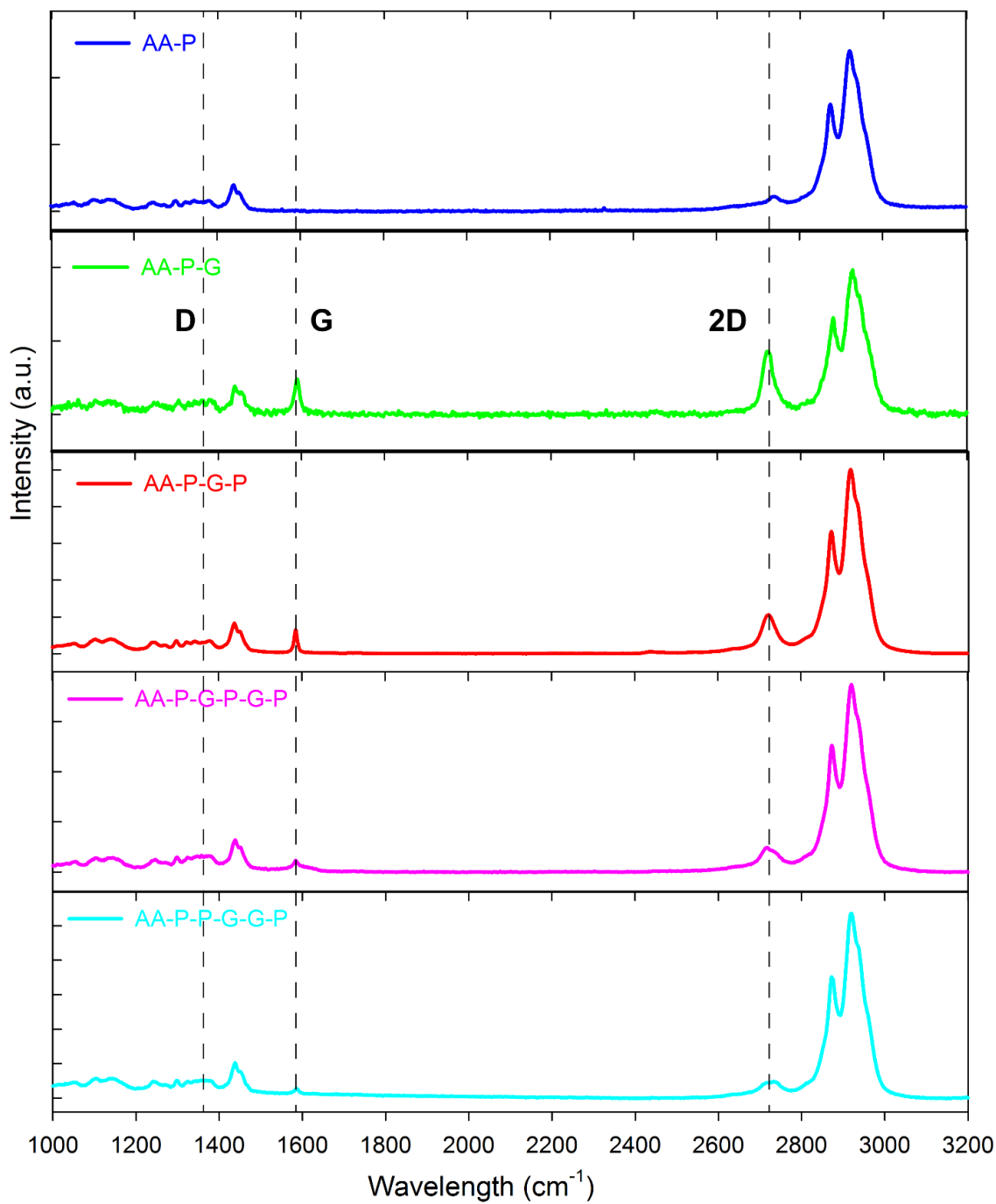

Fig. S5. Raman spectra of the samples after corrosion tests for AA-P (blue), AA-P-G (green), AA-P-G-P (red), AA-P-G-P-G-P (pink) and AA-P-P-G-G-P (cyan), respectively. Vertical dashed lines are used to highlight the characteristic D, G and 2D peaks for graphene.

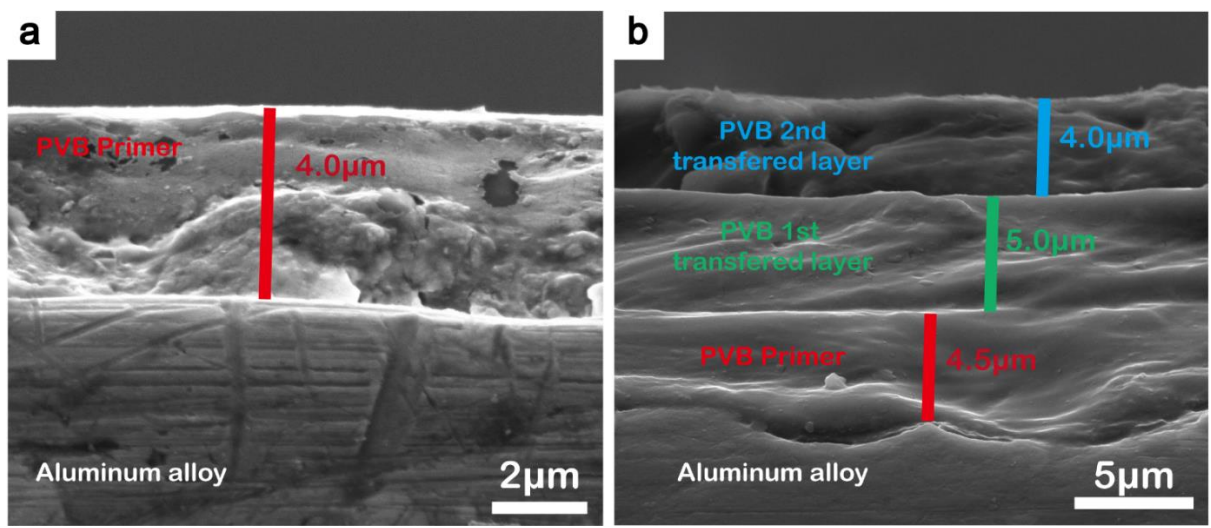

Fig. S6. Cross-section SEM images of (a) spin coated PVB primer on AA substrate and (b) two PVB layers transferred on a PVB primer coated AA substrate. When the PVB layer is either directly spin coated on AA or spin coated on copper and then transferred onto AA, it has a thickness of 4.5±0.5 µm.
7

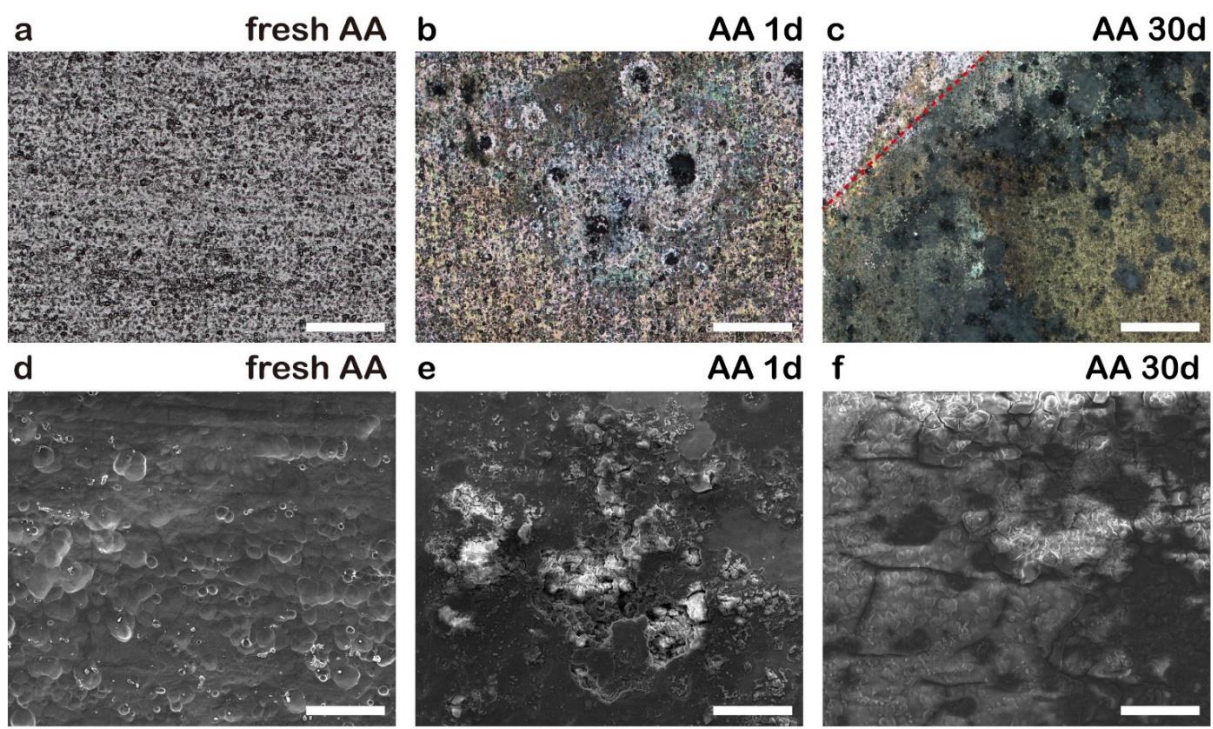

Fig. S7. (a,b,c) Optical images and (d,e,f) SEM images of (a,d) fresh AA, (b,e) AA at 1 day and (c,f) 30 days of immersion in 3.5 wt% NaCl solution. Red dashed line in (c) is to highlight the edge of the O-ring of the corrosion cell. Scale bars are 500 μm in (a,b,c) and 50 μm in (d,e,f).

Localised corrosion or pitting corrosion can be clearly observed on AA after 1 day of immersion in 3.5 wt% NaCl from (b) and (e). After long-term immersion for 30 days, AA is heavily corroded with corrosion products fully covered on its surface, as seen from (c) and (f).

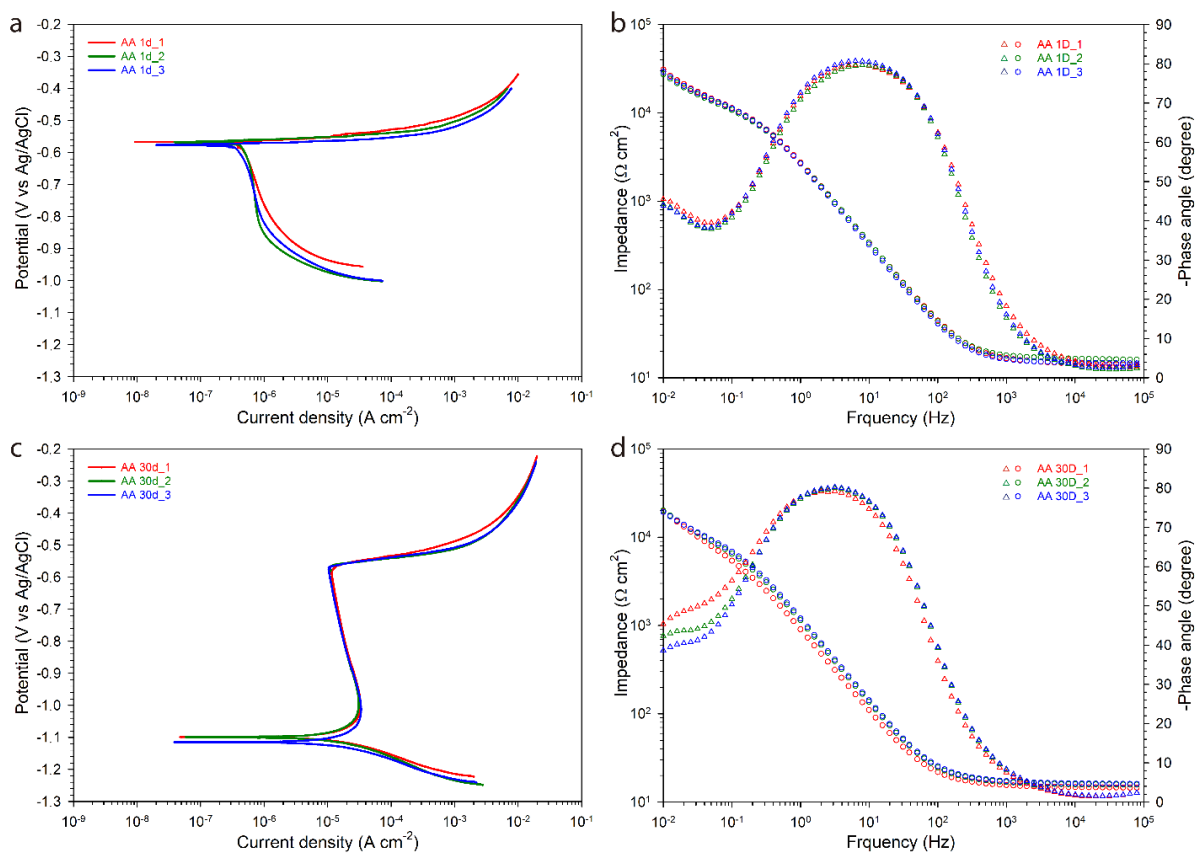

Fig. S8. (a,c) Potentiodynamic polarization curves and (b,d) electrochemical impedance spectra of uncoated AA after (a,b) 1 day and (c,d) 30 days of immersion in 3.5 wt% NaCl solution. For impedance spectra, circles and triangles are data for impedance module and phase angle, respectively.

Additionally, we have observed that results from both measurements for AA show no significant difference between 30 and 60 days of immersion, suggesting that the corrosion current density and impedance of AA are not significantly changed after 30 days of immersion.

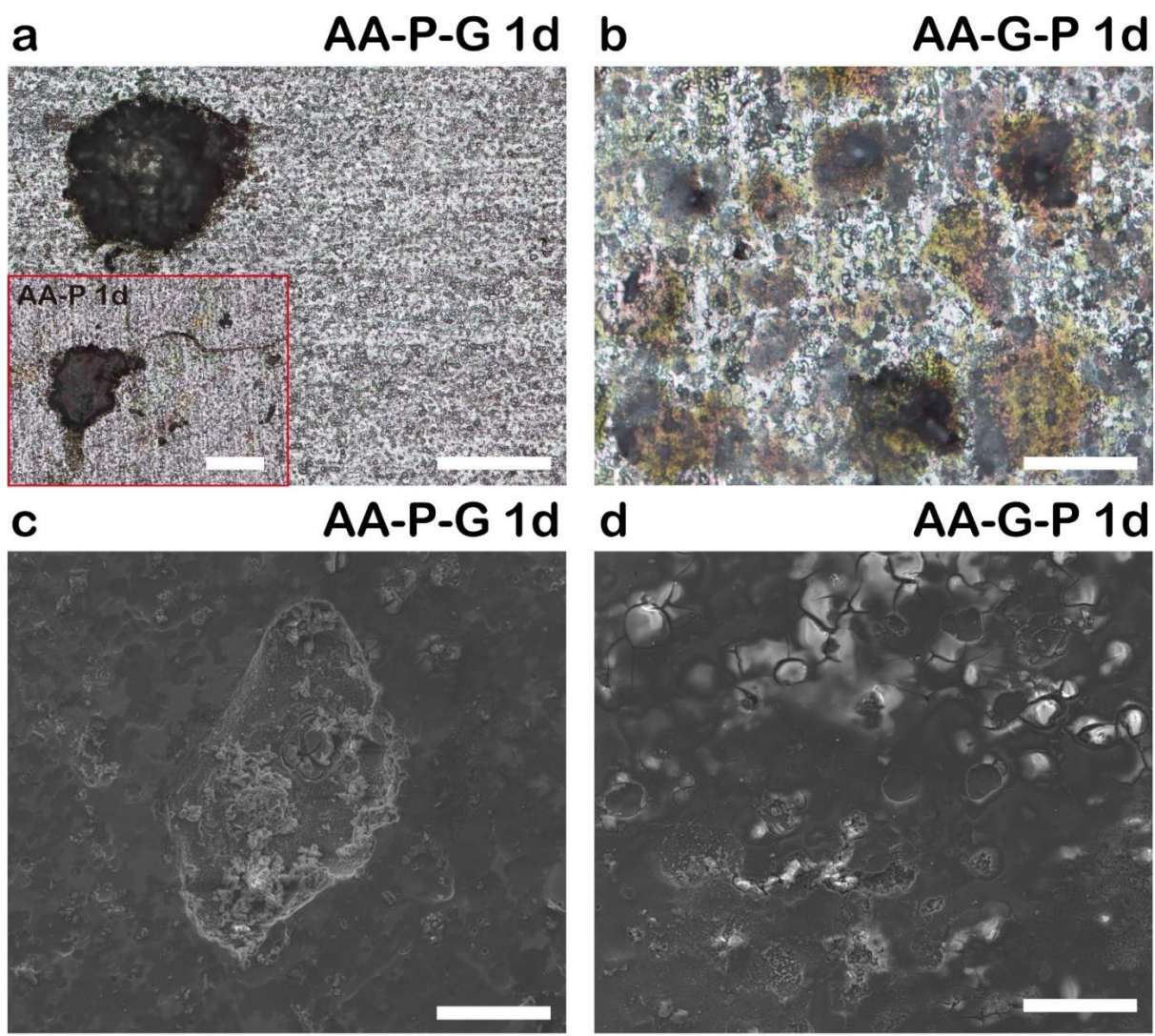

Fig. S9. (a,b) Optical images and (c,d) SEM images of (a,c) AA-P-G and (b,d) AA-G-P after 1 day of immersion in 3.5 wt% NaCl solution. Inset image in (a) presents an optical image of AA-P sample after 1 day of immersion. Scale bars are 500 μm in (a,b) and 50 μm in (c,d).

Localized corrosion can be clearly observed on the AA-P-G sample, as seen from (a, c). However, we notice that AA-P reference sample showed a greater number of dark pits, over a larger area, after 1 day of immersion (see inset of a), while AA-P-G sample corroded in only at a few local spots. This suggest that graphene layer on PVB can provide enhanced barrier performance.

Comparable or accelerated corrosion is observed for AA-G-P sample (c, d) respect to bare AA after 1 day of immersion, indicating that no benefit on corrosion protection is afforded by graphene when it was in direct contact with the AA surface, due to galvanic corrosion and poor adhesion.

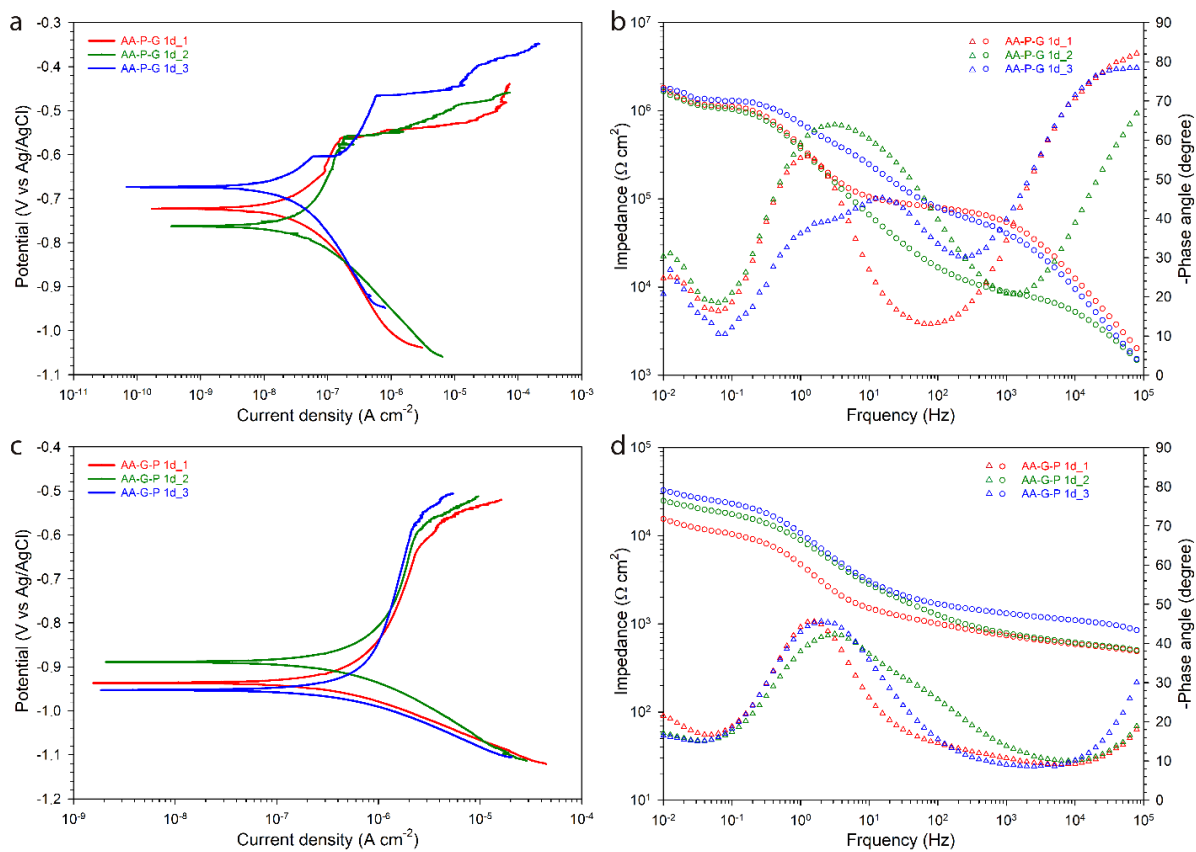

Fig. S10. (a,c) Potentiodynamic polarization curves and (b,d) electrochemical impedance spectroscopy of (a,b) AA-P-G and (c,d) AA-G-P after 1 day of immersion in 3.5 wt% NaCl solution. For impedance spectra, circles and triangles are data for impedance module and phase angle, respectively.

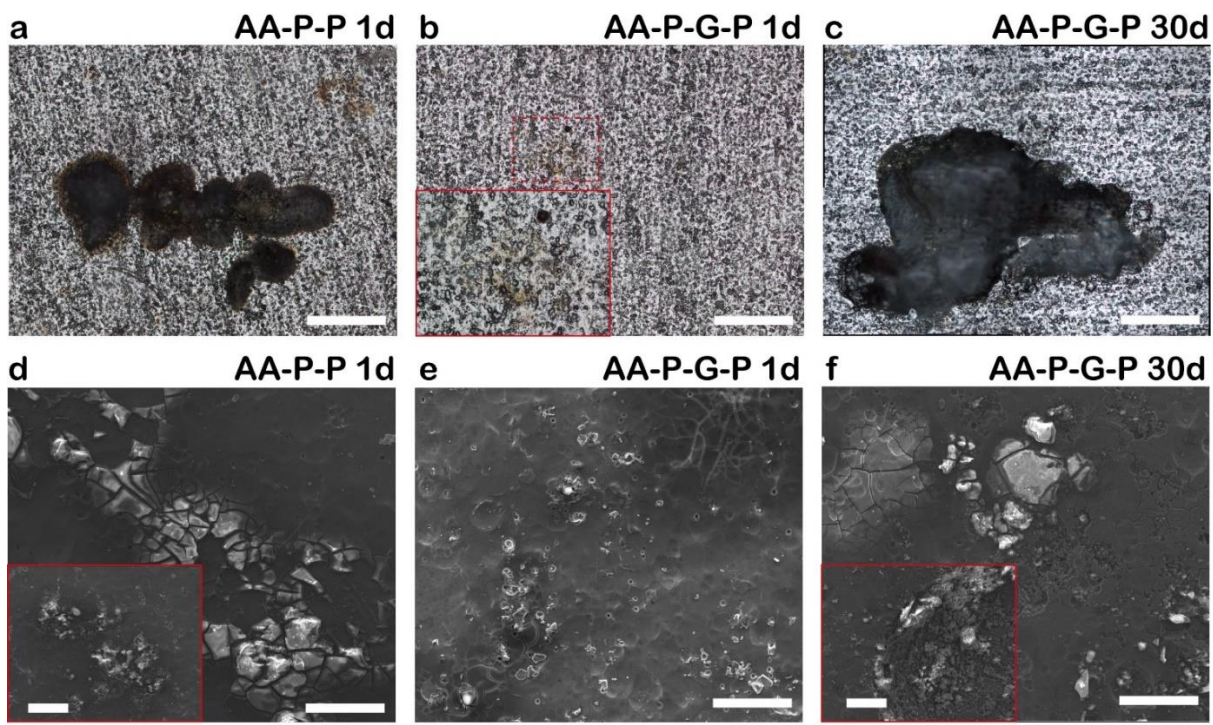

Fig. S11. (a,b,c) Optical images and (d,e,f) SEM images of (a,d) AA-P-P after 1 day and (b,e) AA-P-G-P after 1 day and (c,f) 30 days of immersion in 3.5 wt% NaCl solution. The inset of (b) is a high magnification image of red dashed line highlighted region and insets of (d) and (f) are another typical morphology from other area of the same sample. Scale bars are 500 μm in (a,b,c) and 50 μm in (d,e,f).

System AA-P-P also showed heavy corrosion attack, notably, induced local pitting, as seen from the dark area in (a). SEM observation revealed that a high degree of corrosion developed on AA-P-P, as seen from (d) and the inset image. Much less corrosion attack was observed for AA-P-G-P (b,e) compared with AA-P-P after 1 day of immersion. However, AA-P-G-P showed severe localised corrosion after 30 days of immersion (c,f), where pitting corrosion developed at local defective sites in the graphene. This suggests that AA-P-G-P could offer some effective corrosion protection for AA at short-term immersion of 1 day but not for long-term immersion after 30 days.

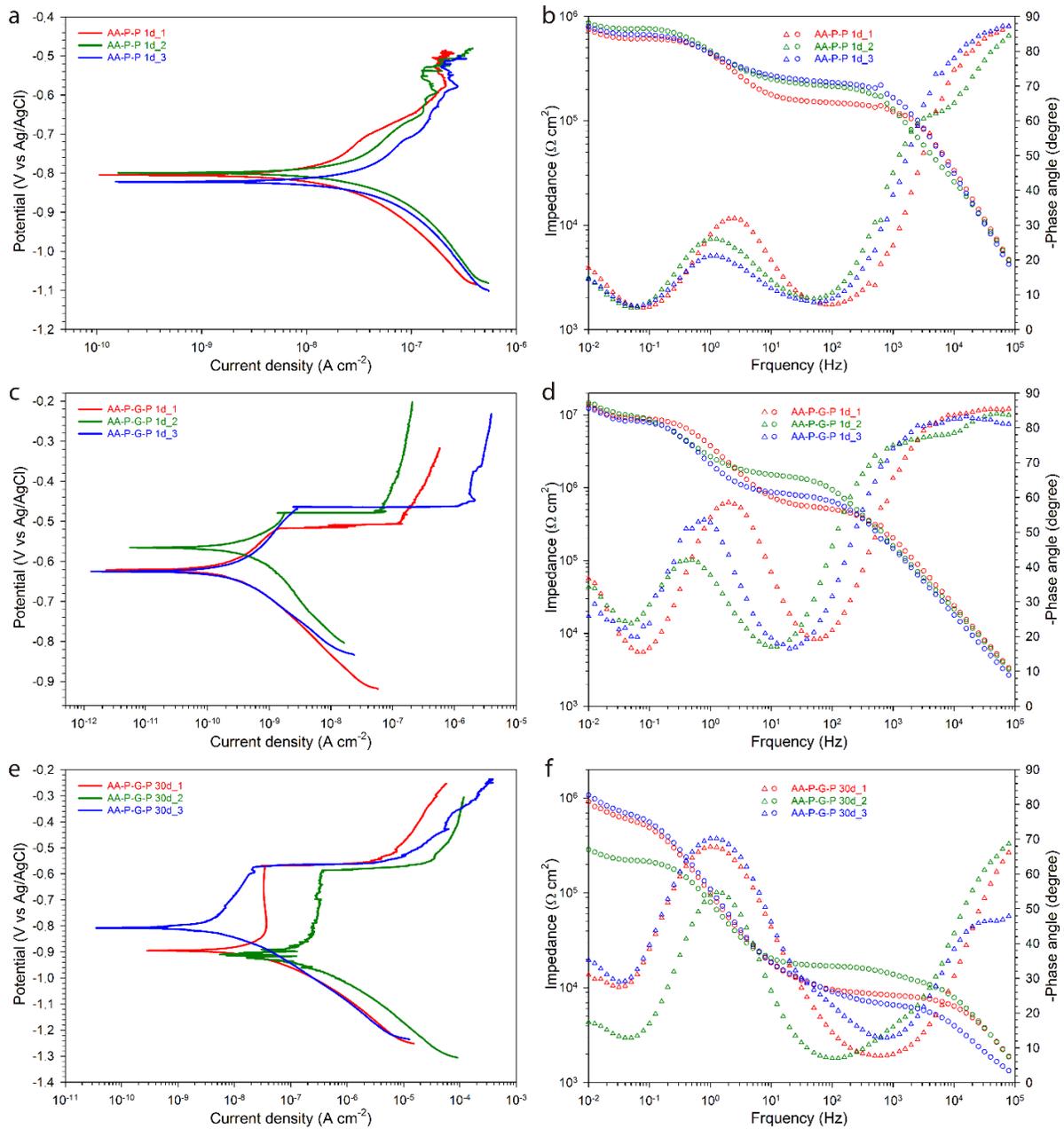

Fig. S12. (a,c,e) Potentiodynamic polarization curves and (b,d,f) electrochemical impedance spectroscopy of (a,b) AA-P-P at 1 day, AA-P-G-P at (c,d) 1 day and (e,f) 30 days of immersion in 3.5 wt% NaCl solution. For impedance spectra, circles and triangles are data for impedance module and phase angle, respectively.

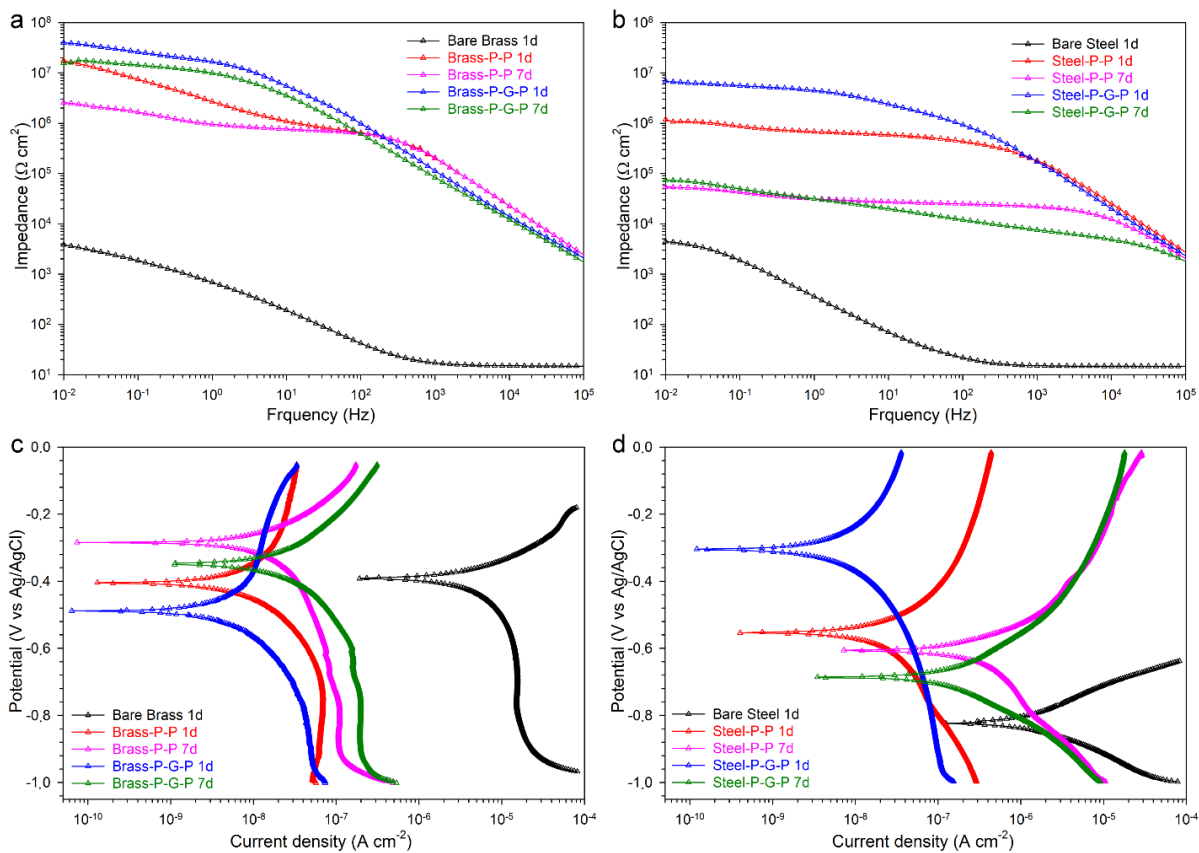

Fig. S13. (a,b) Electrochemical impedance spectroscopy and (c,d) potentiodynamic polarization curves of bare polymer (P-P) and polymer-graphene-polymer (P-G-P) coating on (a,b) carbon steel and (c,d) brass at different days of immersion in 3.5 wt% NaCl solution. Note that these are representive plots of the three reproducible samples.

From both measurements, Metal-P-G-P coatings provide better corrosion protection than Metal-P-P reference coatings for both steel (CXD-2.76.5.90-K, Q-LAB) and brass (Cu63/Zn37, GoodFellow) at 1 day of immersion. However, after 7 days of immersion, there is no significant difference between the performance of P-P and P-G-P coatings when applied on steel. On the other hand, the P-G-P coatings provide better protection than bare P-P ones even after 7 days of immersion when applied on brass.

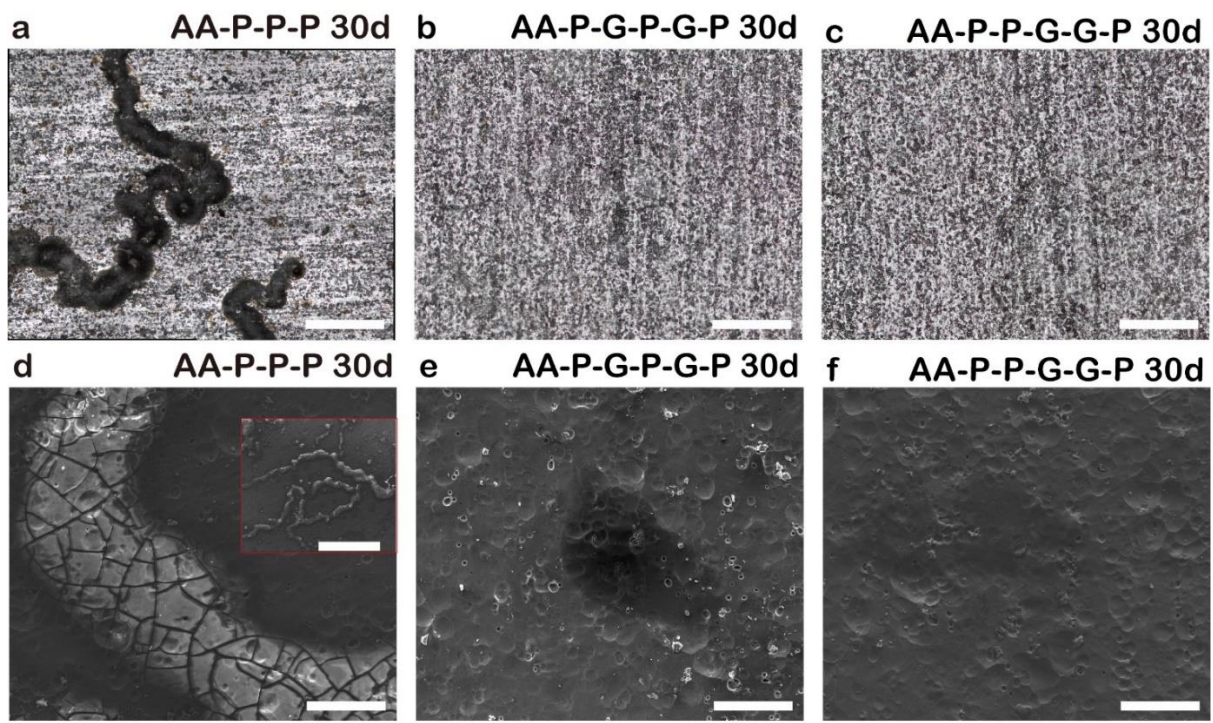

Fig. S14. (a,b,c) Optical images and (d,e,f) SEM images of (a,d) AA-P-P-P, (b,e) AA-P-G-P-G-P and (c,f) AA-P-P-G-G-P after 30 days of immersion in 3.5 wt% NaCl solution. The inset of (d) is a lower magnification image of the same sample. Scale bars are 500 μm in (a,b,c) and 50 μm in (d,e,f). The scale bar in the Inset in (d) is 500 μm.

At 30 days of immersion, bare polymer layer coated AA (AA-P-P-P) showed severe filiform corrosion with heavy corrosion attack at wide range spread across the whole surface, indicating that bare polymer layer could not provide effective corrosion protection for AA after 30 days of immersion. However, when two layers of graphene are sandwiched between three layers of polymer, for both AA-P-G-P-G-P and AA-P-P-G-G-P, after 30 days of immersion, no sign of corrosion was observed from both optical and SEM images.

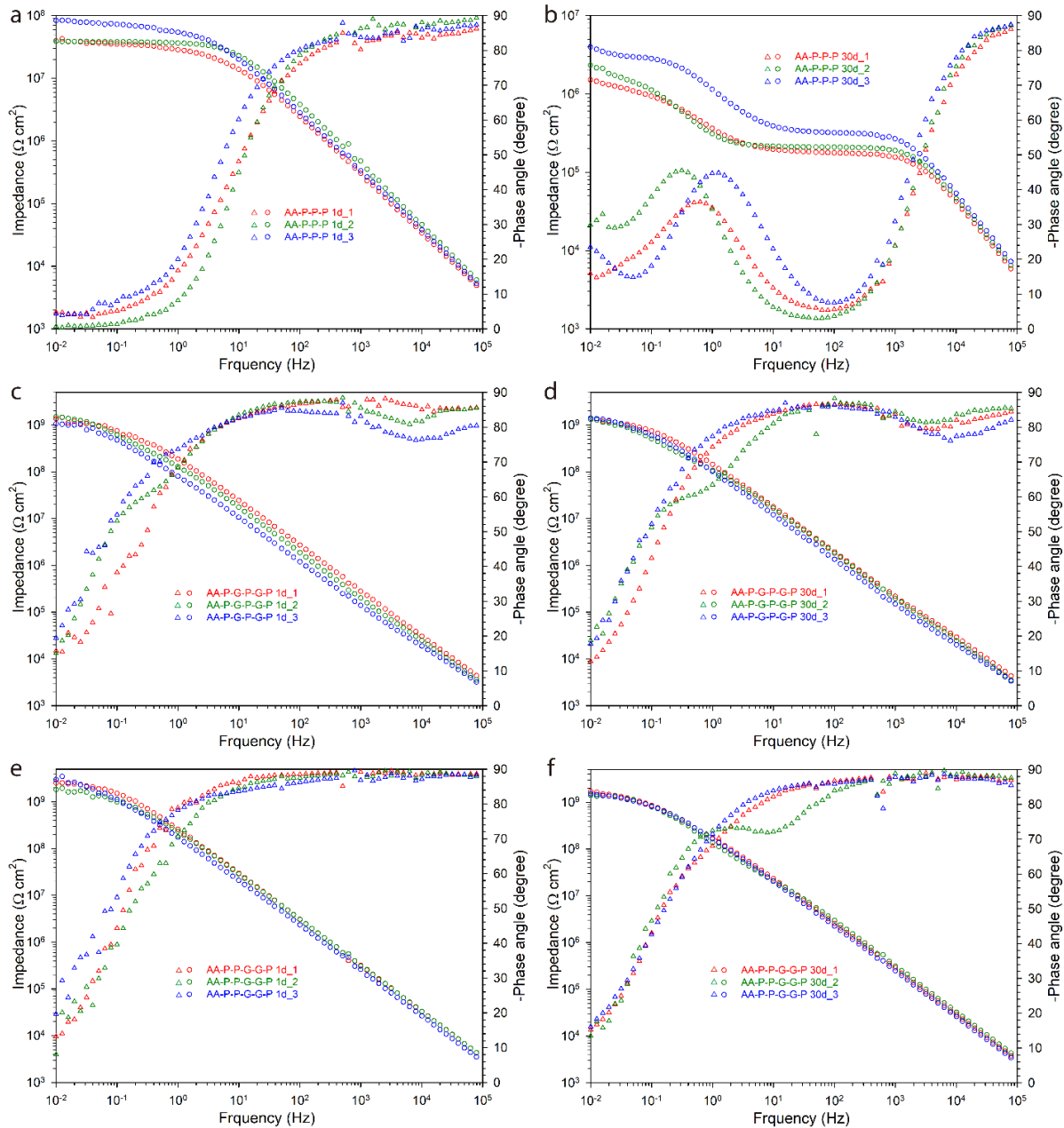

Fig. S15. Electrochemical impedance spectroscopy of (a,b) AA-P-P-P, (c,d) AA-P-G-P-G-P and (e,f) AA-P-P-G-G-G-P after (a,c,e) 1 day and (b,d,f) 30 days of immersion in 3.5 wt% NaCl solution. For impedance spectra, circles and triangles are data for impedance module and phase angle, respectively.

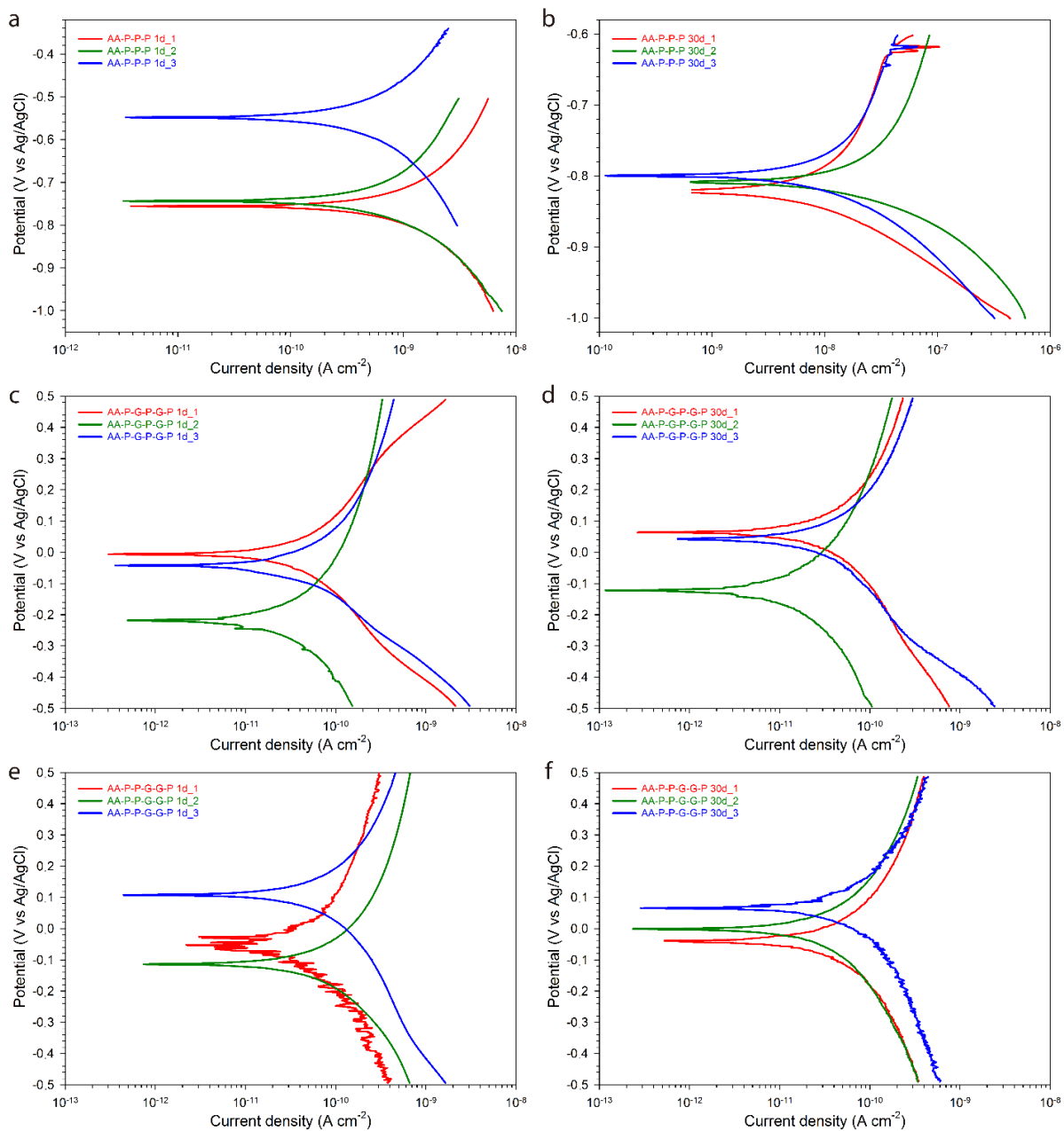

Fig. S16. Potentiodynamic polarization curves of (a,b) AA-P-P-P, (c,d) AA-P-G-P-G-P and (e,f) AA-P-P-G-G-G-P after (a,c,e) 1 day and (b,d,f) 30 days of immersion in 3.5 wt% NaCl solution.

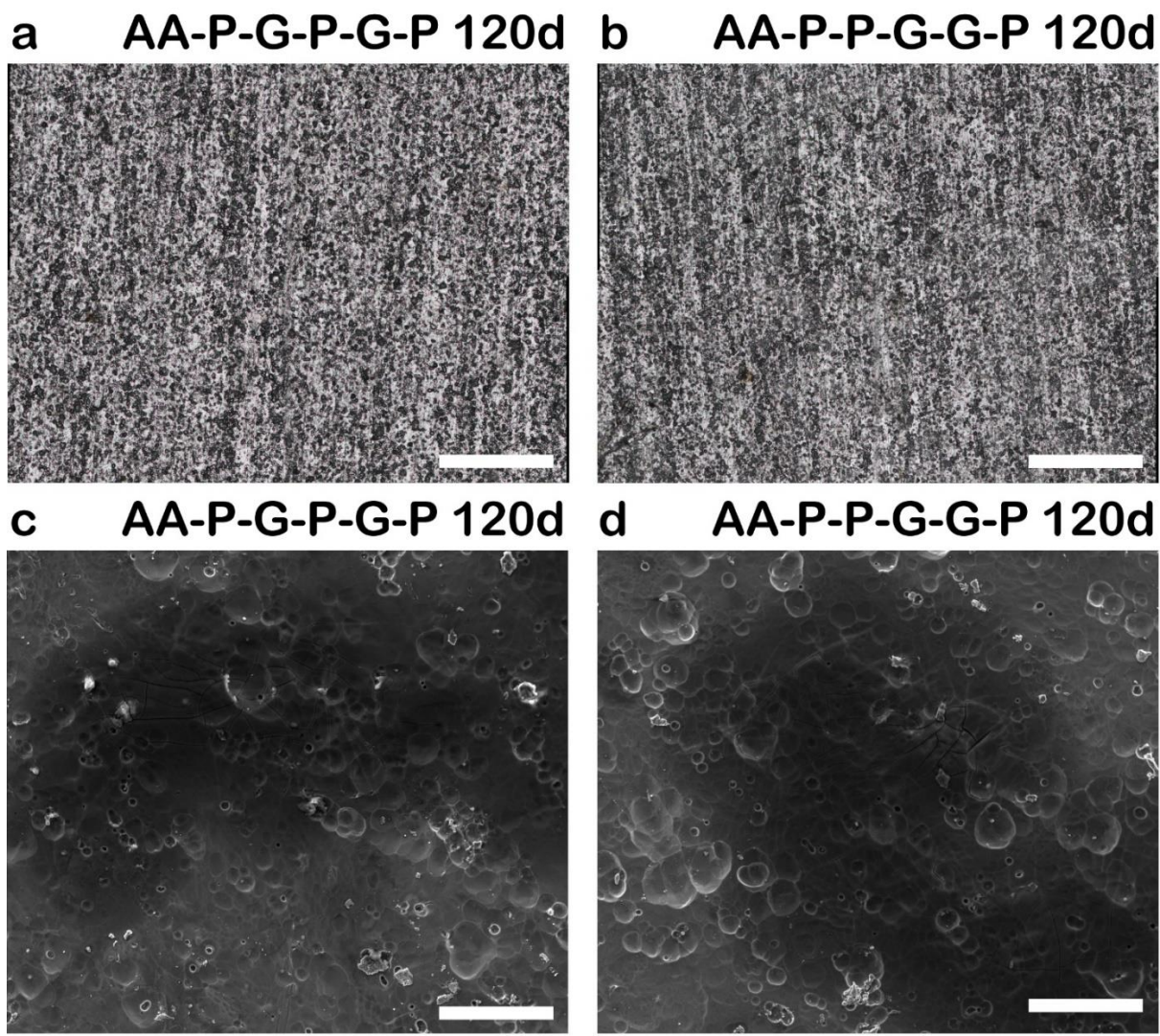

Fig. S17. (a,b) Optical images and (c,d) SEM images of (a,c) AA-P-G-P-G-P and (b,d) AA-P-P-G-G-P after 120 days of immersion in 3.5 wt% NaCl solution. Scale bars are 500 μm in (a,b) and 50 μm in (c,d).

No visible signs of corrosion on AA could be observed from both optical and SEM images for the two samples after 120 days of immersion.

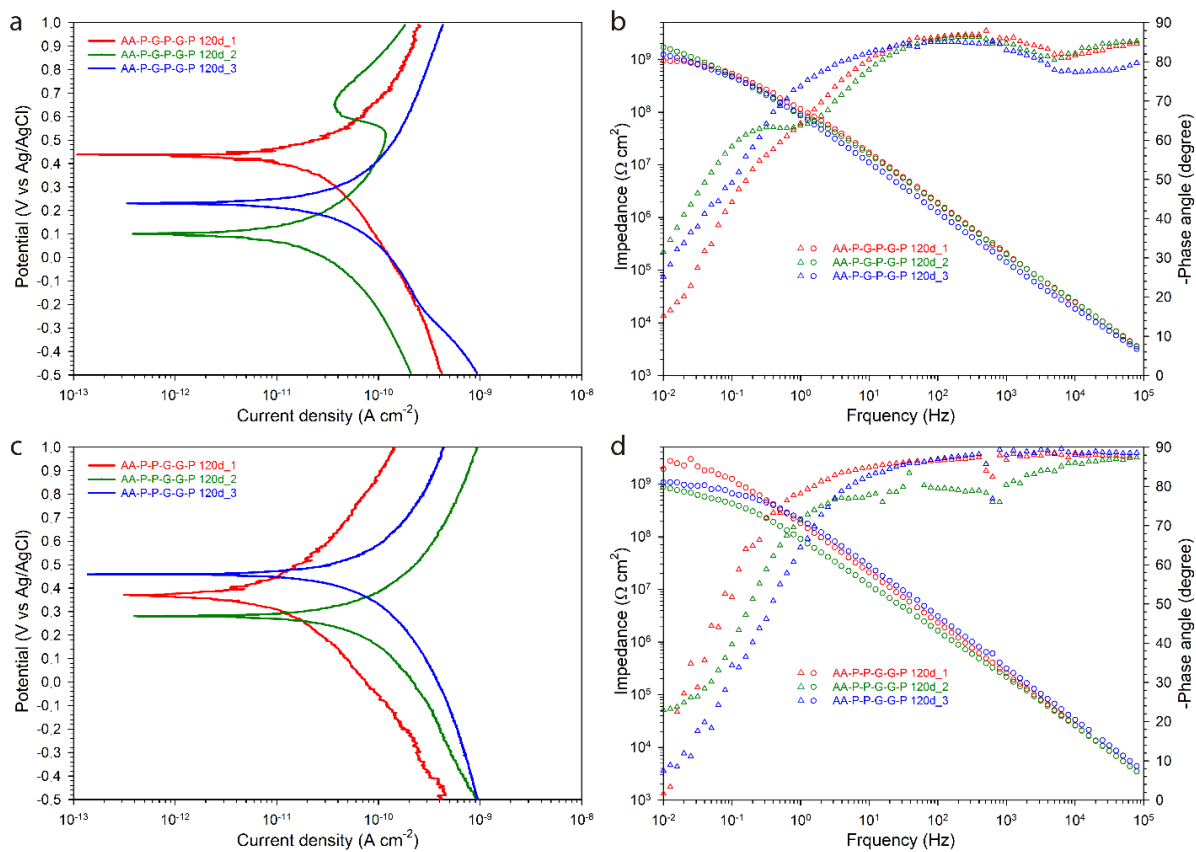

Fig. S18. (a,c) Potentiodynamic polarization curves and (b,d) electrochemical impedance spectroscopy of (a,b) AA-P-G-P-G-P and (c,d) AA-P-P-G-G-G-P after 120 days of immersion in 3.5 wt% NaCl solution. For impedance spectra, circles and triangles are data for impedance module and phase angle, respectively.

| Sample | Open circuit potential OCP (mV vs Ag/AgCl) | | | | Low frequency impedance $\lvert Z \rvert_{0.01Hz}$ (MOhms cm$^2$) | | | |
|---|---|---|---|---|---|---|---|---|
| | 1 | 2 | 3 | average | 1 | 2 | 3 | average |
| AA 1d | -623 | -706 | -697 | -675 | 0.03 | 0.03 | 0.03 | 0.03 |
| AA 30d | -732 | -722 | -738 | -731 | 0.02 | 0.02 | 0.02 | 0.02 |
| AA-G-P 1d | -817 | -825 | -810 | -817 | 0.02 | 0.02 | 0.03 | 0.02 |
| AA-P-G 1d | -591 | -672 | -693 | -652 | 1.82 | 1.67 | 1.75 | 1.75 |
| AA-P-P 1d | -820 | -780 | -797 | -799 | 0.74 | 0.85 | 0.80 | 0.80 |
| AA-P-G-P 1d | -635 | -621 | -657 | -638 | 13.8 | 14.5 | 12.3 | 13.5 |
| AA-P-G-P 30d | -751 | -797 | -732 | -760 | 0.92 | 0.29 | 1.08 | 0.76 |
| AA-P-P-P 1d | -750 | -486 | -501 | -579 | 39 | 40 | 83 | 54 |
| AA-P-P-P 30d | -787 | -758 | -747 | -764 | 1.5 | 2.3 | 3.9 | 2.6 |
| AA-P-G-P-G-P 1d | 870 | 694 | 618 | 727 | 1364 | 1505 | 1077 | 1315 |
| AA-P-G-P-G-P 30d | 731 | 660 | 771 | 721 | 1319 | 1318 | 1408 | 1348 |
| AA-P-G-P-G-P 120d | 587 | 727 | 676 | 663 | 952 | 1709 | 1240 | 1300 |
| AA-P-P-G-G-P 1d | 746 | 487 | 686 | 640 | 2551 | 1830 | 2998 | 2460 |
| AA-P-P-G-G-P 30d | 230 | 448 | 405 | 361 | 1720 | 1408 | 1521 | 1550 |
| AA-P-P-G-G-P 120d | 401 | 204 | 337 | 314 | 1928 | 888 | 1073 | 1296 |

Table S1. Open circuit potential (OCP) and low frequency impedance ($\lvert Z \rvert_{0.01Hz}$) of all samples in this work after immersion times in 3.5 wt% NaCl solution. Values in this table are individual and average values of three samples listed with their corresponding electrochemical impedance spectroscopy.

| Sample | Corrosion potential $E_{corr}$ (mV vs Ag/AgCl) | | | | Corrosion current density $i_{corr}$ (nA cm$^{-2}$) | | | | Corrosion rate CR (μm/year) |
|---|---|---|---|---|---|---|---|---|---|
| | 1 | 2 | 3 | average | 1 | 2 | 3 | average | average |
| AA 1d | -567 | -567 | -576 | -570 | 425 | 390 | 319 | 378 | 4 |
| AA 30d | -1100 | -1100 | -1110 | -1103 | 17700 | 19800 | 18200 | 18567 | 203 |
| AA-G-P 1d | -936 | -889 | -953 | -926 | 651 | 1020 | 837 | 836 | 9 |
| AA-P-G 1d | -723 | -762 | -674 | -720 | 38 | 75 | 35 | 49 | 0.54 |
| AA-P-P 1d | -802 | -799 | -821 | -807 | 52 | 195 | 206 | 151 | 1.6 |
| AA-P-G-P 1d | -619 | -564 | -627 | -603 | 0.39 | 0.53 | 0.34 | 0.42 | 0.005 |
| AA-P-G-P 30d | -894 | -912 | -800 | -869 | 124 | 182 | 12 | 106 | 1.2 |
| AA-P-P-P 1d | -755 | -736 | -551 | -680 | 1.3 | 1.9 | 3.2 | 2.1 | 0.02 |
| AA-P-P-P 30d | -823 | -808 | -798 | -810 | 15 | 43 | 22 | 27 | 0.29 |
| AA-P-G-P-G-P 1d | -9 | -219 | -59 | -96 | 0.08 | 0.27 | 0.09 | 0.15 | 0.0016 |
| AA-P-G-P-G-P 30d | 18 | -158 | -12 | -51 | 0.13 | 0.04 | 0.28 | 0.15 | 0.0016 |
| AA-P-G-P-G-P 120d | 412 | 120 | 191 | 241 | 0.29 | 0.02 | 0.12 | 0.14 | 0.0016 |
| AA-P-P-G-G-P 1d | -47 | -115 | 32 | -43 | 0.23 | 0.22 | 0.11 | 0.19 | 0.0020 |
| AA-P-P-G-G-P 30d | -55 | -36 | 26 | -22 | 0.29 | 0.16 | 0.22 | 0.22 | 0.0024 |
| AA-P-P-G-G-P 120d | 371 | 234 | 461 | 355 | 0.01 | 0.15 | 0.32 | 0.16 | 0.0017 |

Table S2. Corrosion potential ($E_{corr}$), corrosion current density ($i_{corr}$) and calculated corrosion rate (CR) of all samples in this work after different days of immersion in 3.5% NaCl solution. Note that values in the table are individual and average values of three samples listed with their corresponding polarization curves. Corrosion rate are calculated following Faraday's law through equation of CR=$i_{corr}$*K*$E_w$/d, where $i_{corr}$ is corrosion current density (A/cm$^2$), K is corrosion constant (K=3272 mm A$^{-1}$ cm$^{-1}$ year$^{-1}$), $E_w$ is equivalent weight in (9g for Al), d is density (2.7g/cm$^3$ for Al).

| Reference | Metal substrate | Coating system | Electrolyte | Immersion time | Relative corrosion rate improvement ($CR_{uncoated}/CR_{coated}$) | Relative corrosion impedance improvement ($|Z|_{coated}/|Z|_{uncoated}$) |
|---|---|---|---|---|---|---|
| [1] | Cu | CVD Gr | 0.1 M $Na_2SO_4$ | few hours | 7 | 4 |
| [2] | Cu | CVD Gr | 0.1 M NaCl | 1 hour | 50 | 40 |
| [3] | Cu | CVD Gr + ALD $Al_2O_3$ | 0.1 M $Na_2SO_4$ | 3 hours | 100 | 500 |
| [4] | Ni | Thermal annealing grown Gr | 0.1 M NaCl | few hours | 7 | 2 |
| [5] | Ni-Fe alloy | Laser irradiation grown Gr | 0.6 M NaCl | 1 hour | 9 | 7 |
| [6] | Cu | CVD Gr | 0.1 M NaCl | few hours | 10 | 2 |
| [7] | Cu | Electrochemically deposited graphene | 3.5 wt% NaCl | 1 hour | 18 | 3 |
| [8] | Cu | Polymer/graphene composites | 3.5 wt% NaCl | few hours | 11 | 3 |
| [9] | NdFeB | Electrochemically deposited graphene | 3.5 wt% NaCl | few hours | 2 | 10 |
| [10] | Steel | Electrochemically deposited Ni/graphene | 3.5 wt% NaCl | 5 mins | 2 | 2 |
| [11] | Al alloy | Spin coated graphene | 3.5 wt% NaCl | few hours | 2800 | 10 |
| [12] | Steel | Polymer/graphene composites | 3.5 wt% NaCl | 30 mins | 210 | 370 |
| [13] | Steel | Nanocasted epoxy/graphene composites | 3.5 wt% NaCl | few hours | 70 | 3300 |
| [14] | Steel | Polymer/graphene composites | 3.5 wt% NaCl | 30 mins | 200 | 300 |
| [15] | Cu | Polymer/graphene composites | 3.5 wt% NaCl | 100 hours | 140 | 10 |
| [16] | Steel | Electrochemically deposited graphene | 3.5 wt% NaCl | 30 mins | 2 | 3 |
| [17] | Zn | Electrochemically deposited graphene | 3.5 wt% NaCl | 1 hour | 130 | 3 |
| [18] | Fe | Polymer/graphene composites | 3.5 wt% NaCl | 24 hours | 100 | 80 |
| [19] | Steel | Epoxy/graphene composites | 3.5 wt% NaCl | 96 hours | 120 | 10 |
| [20] | Steel | Silane/graphene composites | 3.5 wt% NaCl | few hours | 2000 | 3000 |
| [21] | Al | Dip coated graphene | 0.5 M NaCl | 30 mins | 1200 | 200 |
| [22] | Cu | Polymer/graphene composites | 3.5 wt% NaCl | 30 mins | 110 | 1400 |
| [23] | Fe | Polymer/graphene composites | 3.5 wt% NaCl | 24 hours | 15 | 10 |
| [24] | Steel | Ceramic/graphene composites | 3.5 wt% NaCl | 5 hours | 500 | 50 |
| [25] | Steel | Polymer/graphene composites | 0.5 wt% NaCl | 1 hour | 35 | 150 |
| [26] | Al alloy | Silane/graphene composites | 3.5 wt% NaCl | 30 mins | 580 | 320 |
| [27] | Al alloy | Silane/graphene composites | 3.5 wt% NaCl | 2 hours | 10 | 6 |
| [28] | Steel | Chitosan/graphene composites | 3.5 wt% NaCl | 6 hours | 20 | 2000 |
| [29] | Mg alloy | Silane/graphene composites | 3.5 wt% NaCl | 80 mins | 80 | 15 |
| This work | Al alloy | Polymer/graphene composites | 3.5 wt% NaCl | 30 days | 127000 | 67000 |

Table S3. Comparison of corrosion protection performance between previously reported graphene based anticorrosive coatings[1-29] and this work. Relative corrosion rate (CR) improvement is calculated from the ratio $CR_{(uncoated)}/CR_{(coated)}$. Relative corrosion impedance (|Z|) improvement is calculated based upon the ratio $|Z|_{(coated)}/|Z|_{(uncoated)}$ at 0.01 Hz.